\documentclass{aa}
\usepackage{color}
\usepackage[export]{adjustbox}

\usepackage{graphicx}
\usepackage{mathtools}

\begin{document}

\title{Measuring the continuum polarization\\ with ESPaDOnS\thanks{Based on observations obtained at the Canada-France-Hawaii Telescope (CFHT) which is operated by the National Research Council of Canada, the Institut National des Sciences de l'Univers of the Centre National de la Recherche Scientique of France, and the University of Hawaii.}}

\author{A. Pereyra\inst{1,2} \and C. V. Rodrigues\inst{1} \and E. Martioli\inst{3}}

\offprints{A. Pereyra, \email{antonio.pereyra@inpe.br,apereyra@igp.gop.pe}}

\institute{
Instituto Nacional de Pesquisas Espaciais/MCT, Avenida dos Astronautas 1758, S\~ao Jos\'e dos Campos, SP, 12227-010, Brazil
\and
Instituto Geof\'isico del Per\'u, \'Area Astronom\'ia, Calle Badajoz 169, Lima, Per\'u
\and
Laborat\'orio Nacional de Astrof\'{\i}sica (LNA/MCTI), Rua Estados Unidos, 154, Itajub\'a, MG, Brazil
}

\date{Received dd-mm-yy / Accepted dd-mm-yy}

\abstract
{}
{Our goal is to test the feasibility of obtaining accurate measurements of the continuum polarization from high-resolution spectra using the spectropolarimetric mode of ESPaDOnS.}
{We used the new pipeline OPERA to reduce recent and archived ESPaDOnS data. Several polarization standard stars and science objects were tested for the linear mode. In addition, the circular mode was tested using several objects from the archive with expected null polarization. Synthetic broad-band polarization was computed from the ESPaDOnS continuum polarization spectra and compared with published values (when available) to quantify the accuracy of the instrument.}
{The continuum linear polarization measured by ESPaDOnS is consistent with broad-band polarimetry measurements available in the literature. The accuracy in the degree of linear polarization is around 0.2$-$0.3\,\% considering the full sample. The accuracy in polarization position angle using the most polarized objects is better than 5$\degr$. Consistent with this, the instrumental polarization computed for the circular continuum polarization is also between 0.2$-$0.3\,\%. Our results suggest that measurements of the continuum polarization using ESPaDOnS are viable and can be used to study many astrophysical objects.} 
{}


\titlerunning{Continuum polarization with ESPaDOnS}
\authorrunning{Pereyra et al.}
\maketitle

\section{Introduction\label{intro}}

ESPaDOnS is a bench-mounted high-resolution echelle spectrograph and spectropolarimeter which was designed to obtain a complete optical spectrum from ultraviolet to near-infrared wavelengths in a single exposure (Donati~\cite{don03}, Manset~$\&$~Donati~\cite{man03}). This is installed at the 3.6m Canada-France-Hawaii Telescope (CFHT) in Hawaii. Three operational modes are available in ESPaDOnS: object spectroscopy, sky and object spectroscopy, and spectropolarimetry. The first mode reaches a resolving power of 80,000, and the others up to 65,000.

Polarimetry with ESPaDOnS\footnote{\tiny{http://www.cfht.hawaii.edu/Instruments/Spectroscopy/Espadons/}} was initially conceived to detect linear and circular polarization along line profiles (see, e.g., Wade et al.~\cite{wad05} and Harrington \& Kuhn~\cite{har08}). The evaluation of the feasibility of measuring the continuum polarization during the earlier ESPaDOnS commissioning was not conclusive. A high crosstalk between the Stokes parameters associated with the ESPaDOnS' optics and losses of light over the entrance of the optical fibers were claimed by the ESPaDOnS team as factors that would play against reliable continuum polarization measurements\footnote{\tiny{http://www.cfht.hawaii.edu/Instruments/Spectroscopy/Espadons/ ContiPolar/}}. In consequence, the Libre-ESpRIT pipeline (Donati et al.~\cite{don97}), which has been routinely applied to ESPaDOnS data over the last years, gave priority to the line effect detection offering as default spectropolarimetry with the continuum removed.

We present new results on continuum polarization measurements using ESPaDOnS. A sample of objects, including polarimetric standard stars and science objects with previously measured polarization, was used for this purpose. The new reduction software for ESPaDOnS data, OPERA\footnote{Open-source Pipeline for Espadons Reduction and Analysis (OPERA), is an Open Source software reduction pipeline for the ESPaDOnS spectro-polarimeter at the Canada France Hawaii Telescope.} (Martioli~et~al.~\cite{mar12}), was used in these tests. This pipeline offers the possibility of computing the continuum polarization for ESPaDOnS data.

In Sect.~\ref{instrum} we investigate possible sources of instrumental polarization in the ESPaDOnS polarimetric unit and we present simulations to assess this question. Section~\ref{data} describes the data used in this work including archived and original measurements. The reduction steps processed by OPERA are explained in Sect.~\ref{reduct}. The mathematical procedures used by OPERA to compute the Stokes parameters are shown in Sect.~\ref{measure}. The results of the continuum polarimetry are presented in Sect.~\ref{results} along with the computed synthetic broad-band polarimetry for the sample. A comparison between the different methods implemented by OPERA to compute the polarization is also shown. The perspectives of scientific cases that will benefit from this study are shown in Sect.~\ref{perps}. The conclusions are summarized in Sect.~\ref{concl}.

\begin{table*}
\caption{Log of observations and archived data.}            
\label{tablog}                          
\begin{tabular}{l l c l c c c c c c c}        
\hline\hline
\\[-1ex]                
Object & Other &   \textit{V} & Date & Proposal  & Type &    Observ.      & Detector &  IT$^{b}$ & S/N$^{c}$ & seeing \\  
       & name  &   (mag)      &      & ID$^{a}$  &      &    mode         &          &     (sec) &   &  (\arcsec) \\[1ex]
\hline
\\[-1ex]
\multicolumn{11}{c}{linear polarization}
\\[1ex]
\hline
\\[-1ex]
\object{HD~144432}$^{d}$ & \object{PDS~078}   & 8.2 & 2009-Feb-14 & 09AH10 & HeAeBe & Fast   & EEV1  &  360  & 70  & $-$  \\
\object{HD~150193}$^{d}$ &                    & 8.9 & 2009-May-07 & 09AH10 & HeAeBe & Fast   & EEV1  &  600  & 128 & $-$   \\ 
\object{PDS~069}~N$^{e}$ & \object{Hen~3-949} & 9.7 & 2011-Mar-16 & 11AB04 & HeAeBe & Fast   & Olapa &  650  & 84 & 0.66   \\ 
\object{PDS~395}         & \object{HD~139614} & 8.2 & 2011-Jul-03 & 11AB04 & HeAeBe & Fast   & Olapa &  537  & 245 & 0.60 \\ 
\object{PDS~545}         & \object{HD~174571} & 9.4 & 2011-Jul-05 & 11AB04 & HeAeBe & Fast   & Olapa &  463  & 212 & 0.75 \\    
\object{HD~144668}       & \object{V856~Sco}  & 7.0 & 2011-Jul-10 & 11AB04 & HeAeBe & Fast   & Olapa &  145  & 222 & 0.75  \\  
\object{PDS~080}         & \object{HD~145718} & 8.9 & 2011-Jul-14 & 11AB04 & HeAeBe & Fast   & Olapa &  463  & 175 & 0.52 \\         
\object{PDS~225}         & \object{HD~50083}  & 6.9 & 2011-Nov-13 & 11BB05 & HeAeBe & Normal & Olapa &  336  & 298 & 0.80 \\ 
\object{HD~76543}~S$^{f}$ &                   & 8.0 & 2012-Jan-05 & 11BB05 & HeAeBe & Normal & Olapa &  952  & 267 & 0.65  \\         
\object{PDS~021}         & \object{V791~Mon} & 10.4  & 2012-Jan-08 & 11BB05 & HeAeBe & Normal & Olapa &  1740 & 203 & 0.53 \\         
\object{PDS~281}         & \object{SAO~220669} & 8.9 & 2012-Jan-11 & 11BB05 & HeAeBe & Normal & Olapa &  1232 & 301  & 0.60 \\ 
\object{73~Cet}          &                    & 4.3 & 2012-Jan-10 & 11BE95 & std. unpol. & Normal & Olapa &  15 & 267 &  0.60 \\  
\object{HD~19820}        &                    & 7.2 & 2012-Jan-10 & 11BE95 & std. pol.   & Normal & Olapa &  60   & 159 &  0.60  \\  
\hline
\\[-1ex]
\multicolumn{11}{c}{circular polarization}
\\[1ex]
\hline  
\\[-1ex]
\object{HD~197770} &        & 6.3    & 2008-Jul-22 & 08AO03 & ISM        & Normal  & EEV1  & 760  & $-$ & $-$   \\ 
\object{HD~194279} &        & 7.0    & 2008-Jul-21 & 08AO03 & ISM        & Normal  & EEV1  & 730  & $-$ & $-$  \\
                   &        &        & 2008-Jul-25 & 08AO03 & ISM        & Normal  & EEV1  & 730  & $-$ & $-$   \\
\object{HD~208057} & 16~Peg & 5.1    & 2009-Jul-14 & 09AP13 & Be pole-on & Normal  & EEV1  & 200  & $-$ & $-$   \\
                   &        &        & 2009-Jul-15 & 09AP13 & Be pole-on & Normal  & EEV1  & 200  & $-$ & $-$   \\
\object{HD~175362} &        & 5.4    & 2009-Jul-12 & 09AP13 & non-mag.   & Normal  & EEV1  & 120  & $-$ & $-$   \\
\object{HD~147084} &        & 4.6    & 2006-Jun-12 & 06AD04 & std. pol.  & Normal  & EEV1  & 20   & $-$ & $-$   \\
\object{HD~198478} &        & 4.9    & 2008-Jul-27 & 08BP14 & std. pol.  & Normal  & EEV1  & 60   & $-$ & $-$   \\    
\object{HD~204827} &        & 7.9    & 2011-Jun-14 & 11AP14 & std. pol.  & Normal  & Olapa & 900  & $-$ & $-$   \\
\object{HD~48915}  & Sirius & $-$1.5 & 2011-Feb-19 & 11AE96 & short IT   & Fast    & Olapa & 1    & $-$ & $-$   \\    
                   &        &        & 2011-Feb-19 & 11AE96 & short IT   & Fast    & Olapa & 0.8  & $-$ & $-$   \\    
                   &        &        & 2011-Feb-19 & 11AE96 & short IT   & Fast    & Olapa & 0.6  & $-$ & $-$   \\    
\object{HD~62509}  & Pollux & 1.1    & 2008-Oct-19 & 08BC10 & short IT   & Fast    & EEV1  & 6    & $-$ & $-$   \\
\object{HD~89485}  &        & 4.2    & 2008-Oct-19 & 08BC10 & short IT   & Fast    & EEV1  & 12   & $-$ & $-$   \\
\object{HD~172167} & Vega   & 0.0    & 2011-Jul-15 & 11AE90 & short IT   & Normal  & Olapa & 1    & $-$ & $-$   \\
                   &        &        & 2011-Jul-15 & 11AE90 & short IT   & Normal  & Olapa & 2    & $-$ & $-$  \\     
                   &        &        & 2011-Jul-15 & 11AE90 & short IT   & Normal  & Olapa & 4    & $-$ & $-$   \\            
\hline    
\end{tabular}
\\
\\
$^{a}$ PI names: 08BP14, 08BC10, 09AP13, 11AP14 - G. Wade; 09AH10 - D. M. Harrington; 11AB04 and 11BB05 - A. Pereyra; 06AD04, 11BE95, 11AE96 - N. Manset; 08AO03 - N. Cox; 11AE90 - E. Martioli.\\
$^{b}$ Integration time by individual frame.\\
$^{c}$ Signal-to-noise ratio \textit{per} CCD pixel by single frame for order no.~35 as appeared in headers generated by Libre-ESpRIT.\\
$^{d}$ From CFHT archive.\\
$^{e}$ Northern component.\\
$^{f}$ Southern component.\\
\end{table*}

\begin{table*}[!ht]
\caption{Synthetic optical broad-band polarimetry using ESPaDOnS data and literature values.}            
\label{tabres}                          
\begin{tabular}{l c c c c c c c}        
\hline\hline
\\[-1ex]                
Object & band &  & \multicolumn{2}{c}{ESPaDOnS} &  & \multicolumn{2}{c}{Literature$^{a}$}\\
\cline{4-5}  \cline{7-8}\\
       &        &  & \textit{P}   & PA & &  \textit{P}   & PA \\
       &        &  &  ($\%$)   & ($\degr$) & & ($\%$)    & ($\degr$) \\
\hline\\[-1ex]
\object{HD~144432} & \textit{V} & & 0.27 \;(0.02) & 33.9  \;(2.5)  & & 0.37 \;(0.06) & 14.5  \\
          & \textit{R} & & 0.27 \;(0.03) & 25.7  \;(2.8)  & &             &       \\
          & \textit{I} & & 0.25 \;(0.02) & 16.0  \;(2.7)  & &             &       \\
\object{HD~150193} & \textit{V} & & 5.12 \;(0.05) & 56.1  \;(0.3)  & & 4.78 \;(0.11) & 56.7  \\
          & \textit{R} & & 5.14 \;(0.07) & 56.2  \;(0.4)  & &             &       \\
          & \textit{I} & & 4.83 \;(0.13) & 56.2  \;(0.8)  & &             &       \\
\object{PDS~069}~N  & \textit{V} & & 0.48 \;(0.05) & 153.2 \;(3.3)  & & 0.68 \;(0.04) & 147.0 \\
          & \textit{R} & & 0.65 \;(0.05) & 149.2 \;(2.0)  & &             &       \\
          & \textit{I} & & 0.70 \;(0.03) & 149.1 \;(1.4)  & &             &       \\
\object{PDS~395}   & \textit{V} & & 0.04 \;(0.03) & 96.2  \;(24.2) & & 0.06 \;(0.11) & 147.6  \\
          & \textit{R} & & 0.04 \;(0.02) & 121.4 \;(19.1) & &             &       \\
          & \textit{I} & & 0.04 \;(0.03) & 131.2 \;(21.3) & &            &       \\
\object{PDS~545}   & \textit{V} & & 2.69 \;(0.02) & 81.7  \;(0.2)  & & 3.04 \;(0.03) & 78.6  \\
          & \textit{R} & & 2.56 \;(0.08) & 81.7  \;(0.9)  & &            &      \\
          & \textit{I} & & 2.23 \;(0.10) & 81.5  \;(1.3)  & &            &       \\
\object{HD~144668}$^{b}$ & \textit{V} & & 0.77 \;(0.02) & 16.2  \;(0.9)  & & 0.58 \;(0.02) & 166.5  \\
          & \textit{R} & & 0.81 \;(0.03) & 16.2  \;(0.9)  & &            &        \\
          & \textit{I} & & 0.79 \;(0.04) & 17.5  \;(1.3)  & &            &        \\
\object{PDS~080}   & \textit{V} & & 0.22 \;(0.02) & 82.4  \;(2.7)  & & 0.04 \;(0.05) & 45.8   \\
          & \textit{R} & & 0.28 \;(0.04) & 87.5  \;(4.6)  & &            &        \\
          & \textit{I} & & 0.44 \;(0.07) & 90.0  \;(4.3)  & &            &        \\
\object{PDS~225}   & \textit{V} & & 1.03 \;(0.03) & 32.9  \;(0.9)  & & 0.97 \;(0.04) & 26.3   \\
          & \textit{R} & & 0.96 \;(0.03) & 31.8  \;(1.0)  & &            &        \\
          & \textit{I} & & 0.84 \;(0.04) & 32.0  \;(1.3)  & &            &        \\
\object{HD~76543}~S & \textit{V} & & 0.54 \;(0.03) & 137.9 \;(1.8)  & & 0.47 \;(0.02) & 127.4  \\
          & \textit{R} & & 0.63 \;(0.04) & 138.3 \;(1.7)  & &            &        \\
          & \textit{I} & & 0.68 \;(0.03) & 138.1 \;(1.4)  & &            &        \\
\object{PDS~021}   & \textit{V} & & 1.44 \;(0.04) & 33.9  \;(0.7)  & & 1.60 \;(0.02) & 29.7  \\
          & \textit{R} & & 1.31 \;(0.06) & 32.7  \;(1.4)  & &            &      \\
          & \textit{I} & & 1.05 \;(0.07) & 32.7  \;(2.0)  & &            &       \\
\object{PDS~281}   & \textit{V} & & 1.17 \;(0.04) & 159.0 \;(1.0)  & & 1.38 \;(0.04) & 160.2 \\
          & \textit{R} & & 1.16 \;(0.04) & 162.8 \;(1.0)  & &            &       \\
          & \textit{I} & & 1.04 \;(0.07) & 166.3 \;(1.8)  & &            &        \\
\object{73~Cet}    & \textit{V} & & 0.46 \;(0.02) & 69.4  \;(1.3)  & & 0.09 \;(0.02) & 114.4  \\
          & \textit{R} & & 0.49 \;(0.02) & 69.8  \;(1.4)  & &             &        \\
          & \textit{I} & & 0.54 \;(0.03) & 71.0  \;(1.5)  & &             &       \\
\object{HD~19820}  & \textit{V} & & 4.61 \;(0.02) & 115.1 \;(0.1)  & & 4.79 \;(0.03) & 114.9  \\
          & \textit{R} & & 4.42 \;(0.09) & 115.1 \;(0.6)  & & 4.53 \;(0.03) & 114.5  \\
          & \textit{I} & & 3.88 \;(0.13) & 115.1 \;(1.0)  & & 4.08 \;(0.03) & 114.5  \\
\\[-1ex]
\hline\\[-1ex]
\end{tabular}
\\
Errors in parenthesis.\\
$^{a}$ All objects from Rodrigues et al.~(\cite{rod09}) except \object{73~Cet} and \object{HD~19820} (from Schmidt~et~al.~\cite{sch92}).\\
$^{b}$ This object presents confirmed variability in polarization (especially in PA). Hutchinson~et~al.~(\cite{hut94}) reported  $P(V) = 1.19\,\%~(0.08\,\%)$ at $\rm{PA} = 13\fdg4~(5\fdg5)$, and Bhatt~(\cite{bha96}) $P(V)= 0.58\,\%~(0.11\,\%)$ at $\rm{PA} = 168\degr~(5\degr)$. Our result is more consistent with  Hutchinson~et~al.~(\cite{hut94}), and Rodrigues et al.~(\cite{rod09}) with Bhatt~(\cite{bha96}).
\end{table*}

\section{Instrumental limitations of ESPaDOnS when measuring continuum polarization\label{instrum}}

Donati et al.~(\cite{don99}) affirmed that the ``MuSiCoS spectropolarimeter can in principle also estimate continuum polarization in stellar spectra. ... The main problem here consists in reducing true and spurious instrumental polarisation to a minimum". ESPaDOnS has a very similar design  to MuSiCoS. Therefore, if the instrumental effects degrading accuracy and precision of polarization measurements using ESPaDOnS are understood, one can efficiently reduce and/or quantify them and perform reliable measurements of continuum polarization using this instrument. In this section we discuss these effects when they originate after light has entered the polarimetric unit. We also present simulations including some of these issues.


\begin{figure}[!h]
\centering
\includegraphics[scale=0.4]{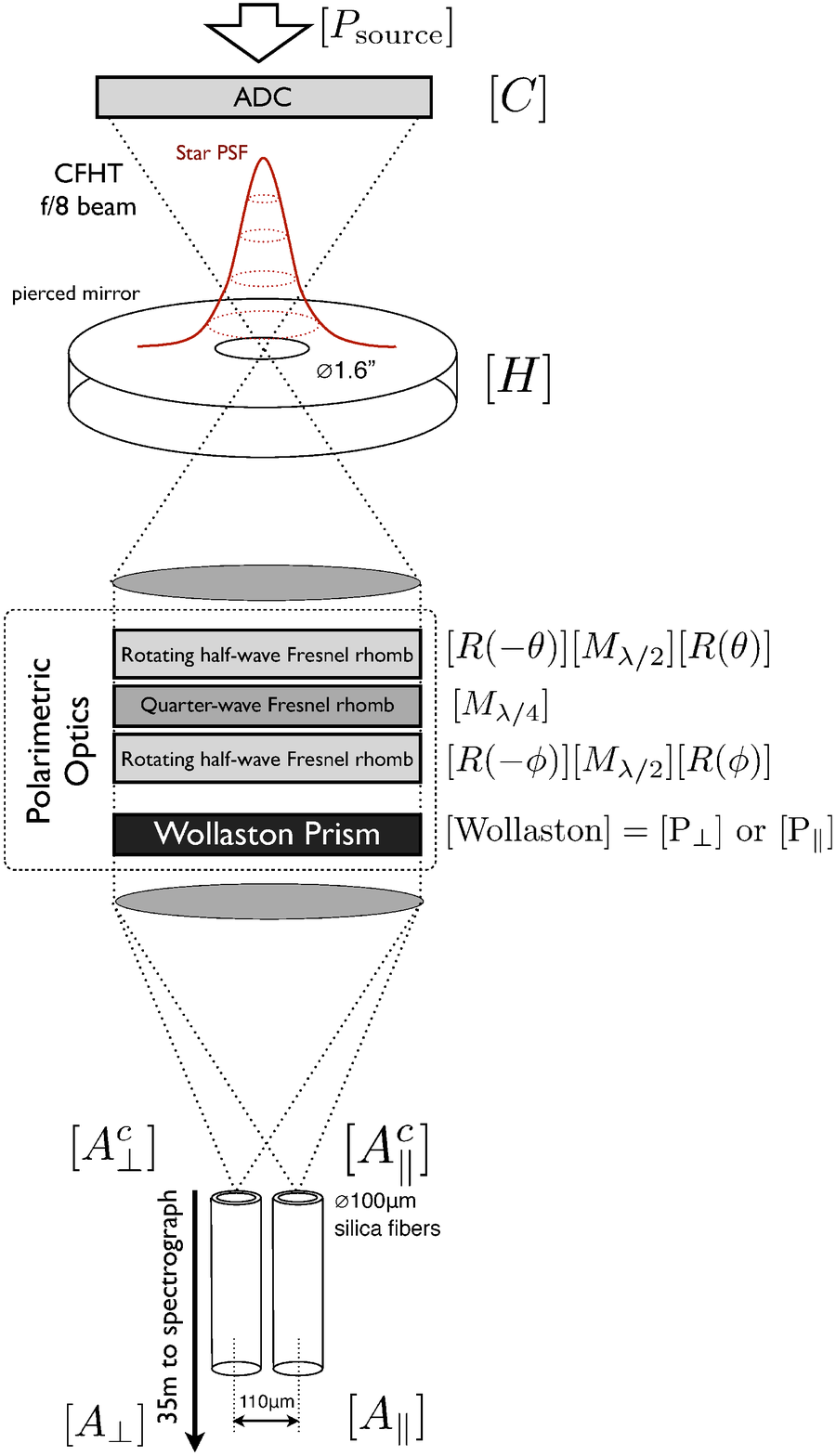}
\caption{Optical path of the polarimetric unit of ESPaDOnS.} 
\label{polopt}
\end{figure}

Figure \ref{polopt} illustrates the optical path of the polarimetric unit of ESPaDOnS, where the elements that may introduce spurious polarization are shown. Figure \ref{polopt} also shows the symbols representing the Mueller matrices associated with each optical element, some of which are used in our simulations.

Mueller matrix calculus can be applied to simulate the net effect of polarization introduced by the optical train in the input light. This set of matrix transformations must be organized in the correct order and then applied to the four-component Stokes vector describing the input light to obtain the output intensity of each measured beam of ESPaDOnS. Adopting the same definitions as in de~la~Chevroti\'ere et al. (\cite{che13}), the simulated output beam is given by $[{\bf S}_{o}] = [I_{o},Q_{o},U_{o},V_{o}]$, which can be calculated as
\begin{equation}
[{\bf S}_{o}] = [P][{\bf S}_{i}], 
\end{equation}
where the input beam is given by $[{\bf S}_{i}] = [I_{i},Q_{i},U_{i},V_{i}]$, and the transformation matrix $[P]$ is given by the product between all Mueller matrices shown in Fig.~\ref{polopt}, i.e.,
\begin{equation}
\begin{multlined}
[P]_{\perp} =  [A_{\perp}][A^{c}_{\perp}][P_{\perp}]R(-\phi)][M_{\lambda/2}][ R(\phi)] \\
\;\;\;\;\;\;\;\; [M_{\lambda/4}][R(-\theta)][M_{\lambda/2}][ R(\theta)] [H][C][P_{\rm source}],
\end{multlined}
\end{equation} 
and
\begin{equation}
\begin{multlined}
[P]_{\parallel} =  [A_{\parallel}][A^{c}_{\parallel}][P_{\parallel}]R(-\phi)] [M_{\lambda/2}][ R(\phi)] \\
\;\;\;\;\;\;\;\; [M_{\lambda/4}][R(-\theta)] [M_{\lambda/2}][ R(\theta)] [H][C][P_{\rm source}], 
\end{multlined}
\end{equation}
where $[P]_{\perp}$ and $[P]_{\parallel}$ are the perpendicular and parallel components.

The first element shown in Figure \ref{polopt} is the atmospheric dispersion corrector (ADC), which is an optical device that may present some stress birefringence in its optical parts, introducing spurious crosstalk polarization. The crosstalk matrix above is represented by $[C]$. Crosstalk is an instrumental artifact that may appear in both continuum and line polarization measurements. It is the conversion between fluxes in different Stokes parameters. Another possible source of crosstalk polarization is the inaccuracy of the positioning of the rotating rhombs. In ESPaDOnS, the total crosstalk between linear and circular polarization has been monitored for several years. It was reduced from 10\,\% to 0.3\,\% in 2009, as decribed in the ESPaDOnS homepage and Barrick et al.~(\cite{bar10}). The present crosstalk represents a very small fraction of a given polarized source, which can be considered negligible in the study presented here. Hence, we are not concerned about crosstalk effects and for the moment we will just ignore the contribution from both $[C]$ and from the positioning inaccuracy in $\theta$ and $\phi$. The interested reader may consult Bagnulo et al. (\cite{bag09,bag12}), de la Chevroti\'ere et al.~(\cite{che13}), or Ilyin~(\cite{ily12}) for a quantitative discussion and procedures to remove crosstalk from spectropolarimetric measurements.

According to the ESPaDOnS homepage\footnote{\tiny{http://www.cfht.hawaii.edu/Instruments/Spectroscopy/Espadons/ Espadons\_FAQ.html}}, there are two main factors that prevent reliable continuum polarization measurements:

\begin{enumerate}

\item Instrumental polarization can occur when the stellar light enters the hole of the pierced mirror, where a non-uniform illumination of the edges of the hole introduces polarization. This effect can be accounted for by matrix $[H]$.  This is expected to depend on the position of the star relative to the center of the entrance hole and on the source extension. As mentioned in the ESPaDOnS homepage this effect can be minimized by either improving guiding or using integration times where random guiding errors can be averaged out. This effect is also reduced under poor seeing conditions, where the illumination function is smoother at the edges, reducing the amplitude of asymmetries caused by decentralization of the source with respect to the hole.  The polarization matrix $[H]$ is somewhat difficult to model, since one needs to know the polarizing properties of the metallic surface of the pierced mirror. For this reason we decided to address this issue by inspecting the dependence between the spurious polarization and seeing conditions from our dataset. This is discussed in Sect.~\ref{results}.
\\
\item The second effect occurs after the beam is split by the Wollaston prism. We note that at this stage the light beam has already passed the polarizer and analyzers and therefore the polarimetry is affected only by Stokes \textit{I}. There are two problems associated with this stage. 
First, each of the two resultant source images reaches a different optical fiber. If each beam is not the same distance from  the fiber center, then the fluxes entering each fiber will be different. This effect is accounted for by a pair of attenuation matrices, $[A^{c}_{\perp}]$ and $[A^{c}_{\parallel}]$. Second, the differences in the transmission of the two fibers will also appear as an instrumental spurious polarization. This effect is accounted for by the attenuation matrices, $[A_{\perp}]$ and $[A_{\parallel}]$, which will introduce fiber transmission losses.

\end{enumerate}

Attenuation matrices are defined by one parameter, $p$, which is a number between $0$ and $1$ that gives the fraction of light intensity that goes through the attenuator. For $[A^{c}_{\perp}]$ and $[A^{c}_{\parallel}]$, $p$ depends on the fraction of light collected by the fiber and should also be a function of guiding errors and seeing conditions. For the sake of simplicity, we assume in the simulations that guiding errors are random and thus are averaged out over an exposure. We also disregard the dependence on atmospheric seeing conditions. According to Sect.~\ref{seeingit}, its impact is negligible in median seeing conditions at the CFHT site (about  0$\farcs$7), given that appropriated integration times are used. We also note that the aperture entrance for ordinary and extraordinary beams is 100\,$\mu$m, which corresponds to 1$\farcs$58, more than twice the average seeing.


For the present simulation we are only concerned about factors $[A_{\perp}]$ and $[A_{\parallel}]$, for which the attenuation parameter $p$ is given by the fiber transmission, $T$. Optical fiber transmissions can vary due to slight differences in the fiber optical path caused by different torsions or temperatures. ESPaDOnS is equipped with fiber scramblers working at 30\,Hz and amplitude of 1\,mm. This minimizes the transmission variations due to torsions.

\begin{figure}
\centering
\includegraphics[scale=1,clip]{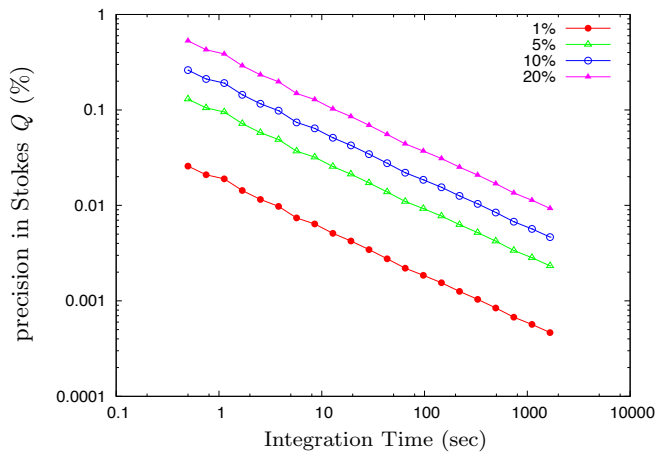}
\caption{Simulations of polarimetric precision in degree of polarization for Stokes \textit{Q} versus integration time. These simulations only consider the effects caused by variability in fiber transmission. These results were obtained from 500 iterations, where we  used four different values for the uncertainty in fiber transmission ($\delta$\textit{T}), 1\,\%, 5\,\%, 10\,\%, and 20\,\%, as indicated in the legend.} 
\label{polpre}
\end{figure}

All the above effects are expected to vary slowly with wavelength. Therefore, if the observer is interested in a small portion of the spectra, for example spectral lines studies, the underlying polarization continuum can be subtracted and any instrumental polarization will also be removed. This is the standard procedure adopted in ESPaDOnS data reduction.

To simplify and isolate the possible sources of errors we have only considered here the effects caused by optical transmission losses in the fibers. In this sense, we consider that a given fiber upon a movement will present a variation in the throughput. If the fiber transmission is given by $T$ then the actual transmission after a given movement is given by $T\pm\delta T$, where $\delta T$ is the square root of the variance of transmission caused by one movement of the fiber, assuming that the fiber transmission changes randomly with each movement. By assuming a normal distribution for the fiber transmission, the amplitude of the variance of transmission after a given number of movements $N$ is given by
\begin{equation}
{\Delta T}^{2}= \frac{{\delta T}^{2}}{N},
\end{equation}
where $N=f_{s}T_{\rm exp}$, with $f_{s}$ being the frequency of movements or the fiber scrambler frequency in Hz and $T_{\rm exp}$ being the integration time in seconds. The final variance for the fiber transmission at the end of an exposure is then given by

\begin{equation}
{\Delta T}^{2}= \frac{{\delta T}^{2}}{f_{s}T_{\rm exp}}.
\label{eq:fibertransvar}
\end{equation}

Equation \ref{eq:fibertransvar} shows that transmission uncertainty is inversely proportional to the square root of integration time. By reducing the uncertainty in fiber transmission there should also be an improvement in the polarimetric precision in what concerns the instrumental polarization introduced by variations in \textit{T}.

To quantify this effect we have run simulations where we adopted a nominal mean fiber transmission of $T=80\,\%$, which is the ESPaDOnS fiber transmission at $\lambda=600$\,nm. The simulations are run for an unpolarized source that enters the optical train. Then we apply the Mueller calculus to estimate the net effect of polarization from all components explained above to obtain the resulting flux for each beam (parallel and perpendicular) in each exposure. This procedure is repeated four times considering the typical number of exposures used by ESPaDOnS to measure each Stokes parameter.

Then the resulting fluxes are used as input to the OPERA pipeline where the polarization is calculated using the same algorithm as in real measurements. The simulations are run for 500 measurements of polarization, where we obtain the mean degree of polarization $<P>$ and the standard deviation $\sigma_{P}$. The latter represents a quantitative measurement of the precision in polarization. We run the same simulations for four different values of transmission uncertainty in one movement of the fiber: $\delta T = $ 1\,\%, 5\,\%, 10\,\%, and 20\,\%. Figure \ref{polpre} shows the results of these simulations. It shows that polarimetric precision improves significantly for longer integration times, where for the highest uncertainty considered in fiber transmission of $\delta T = $ 20\,\%, integration times longer than 10\,sec provide a polarimetric precision better than 0.1\,\%.


\subsection{Comments about 2005 commissioning data}

The ESPaDOnS webpage\footnote{\tiny{http://www.cfht.hawaii.edu/Instruments/Spectroscopy/ Espadons/ContiPolar/}} 
shows measurements of $\gamma$ Cas (HD~5394) obtained in September 2005 as part of a commissioning run to test the stability of continuum polarization measurements with ESPaDOnS. We note that $\gamma$ Cas is not an unpolarized star. It is a Be star and its polarization is around 1\,\% and variable at a level of 0.10$-$0.15\,\% (Piirola~\cite{pii79}). For the spectral range where $\lambda>540$\,nm these data present high levels of instrumental polarization that reach up to 10\,\%. However for the spectral range where $\lambda<540$\,nm this variation remains below $<2$\,\%. These observations have used integration times of 5\,sec.   The levels of instrumental polarization in this experiment are well above the precision in polarization calculated in the simulations discussed in this section, and therefore cannot be explained by transmission losses in the fibers of up to 20\,\%.  However, ESPaDOnS has been through two significant upgrades since these measurements were done, which should be taken into account.

The first upgrade was the replacement of the ADC triplet lenses, which occurred in two stages. The first replacement was installed in June 2006 and the second replacement was installed in October 2008. The old ADC presented high levels of birefringence 
and optical imperfections due to excessive stress in the optical parts. This work is well explained in Barrick~et~al.~(\cite{bar10}).

The second upgrade was the installation of new fibers and connectors, which occurred in June 2006. During the commissioning mentioned above ESPaDOnS had a spare fiber bundle installed, which replaced the original bundle between June 2005 and Feb 2006 (see the ESPaDOnS webpage). This spare bundle exhibited a throughput loss of 0.4\,mag, which represents a change of about 45\,\% in flux.

We have two hypotheses that can explain the high levels of polarization seen in the commissioning data which agrees with both upgrades. The first is that imperfections in the ADC lenses could cause chromatic aberration in the star point spread function (PSF), which could introduce a spectral dependency in the alignment of the star with respect to the entrance hole. This could propagate to a misalignment between the two images produced by the Wollaston prism and the two fiber inputs as well. The ESPaDOnS guider uses a SCHOTT/BG38 low pass filter to select visible light\footnote{\tiny{http://www.ast.obs-mip.fr/projets/espadons/espadons\_new/ guiding.html}}, with cut-off wavelength starting at $\lambda>550$\,nm\footnote{\tiny{http://www.schott.com/advanced\_optics/english/ download/schott-optical-filter-glass-properties-2013-eng.pdf}}.

The detector for guiding is a Finger Lake Instrumentation EEV CCD of type CCD47-10, which has peak efficiency at about $\lambda=$ 510\,nm\footnote{\tiny{http://www.flicamera.com/spec\_sheets/ML4710.pdf}}. Both efficiencies combined give a maximum in about the same spectral range where ESPaDOnS presents best performance in continuum polarimetric precision seen in the commissioning data. This makes us believe that centering works correctly for spectral regions where the guider is most sensitive. Chromatic effects caused by the ADC produces a degradation for other spectral ranges. This  effect has probably been fixed (or minimized) after the ADC was replaced. 

The second hypothesis is that the old fibers presented significantly higher throughput losses ($\delta\textit{T} > 45\,\%$), which combined with the short integration time of 5\,sec could have caused the high levels of spurious polarization seen in those data. In either case the problems have been mitigated and this signature does not seem to be present in recent data, obtained after the ADC and fiber repairs, as will be shown in subsequent sections.

\section{Data description\label{data}}

In this section, we describe the data used to demonstrate the feasibility of measuring continuum polarization using ESPaDOnS. We present the linear and circular polarization data separately.

\subsection{Linear polarization}


The dataset for linear polarization is composed of original observations of Herbig  AeBe (HeAeBe) stars gathered with ESPaDOnS at CFHT on semesters 2011A and 2011B (Proposal IDs 11AB04 and 11BB05). This program was initially proposed to measure the linear polarization across the H$\alpha$ line. The results concerning the H$\alpha$ measurements will be published in a separate paper. A total of nine HeAeBe stars were measured and a log of observations is shown in Table~\ref{tablog}. All the HeAeBe stars have optical linear polarization published (Rodrigues~et~al.~\cite{rod09}) which allows us to do a proper comparison.

In order to test the capability of the OPERA software to reduce archived data of ESPaDOnS and enlarge our sample, we selected a couple of HeAeBe stars from the CFHT database\footnote{\tiny{http://www.cfht.hawaii.edu/ObsInfo/Archive/}} (Proposal ID 09AH10). These stars (\object{HD~144432} and \object{HD~150193}) also have previously published optical polarimetry (Rodrigues~et~al.~\cite{rod09}). Finally, two polarimetric standard stars from the Atlas of HST Polarimetric Calibration Objects (Turnshek et al.~\cite{tur90}) were measured using the same ESPaDOnS configuration in the 2011B semester as part of an independent program (Proposal ID 11BE95). One of them is the unpolarized star \object{73~Cet} and the other one is the polarized star \object{HD~19820}. These additional four objects are included in Table~\ref{tablog}.

Our sample spans \textit{V} magnitude between 4.3 and 10.4\,mag. The archived data (observed in 2009A) and the 2011A HeAeBe stars were observed using the fast readout mode of ESPaDOnS. The rest of our sample (2011B) was observed using the normal readout mode, which includes a slightly larger overhead between individual exposures. The archived data were taken using the old ESPaDOnS detector EEV1 and the rest of the sample with the new and enhanced detector Olapa. The integration time (IT) by individual frame is indicated in Table~\ref{tablog}. Considering that a set of eight frames is needed to characterize completely a linear spectropolarimetric measurement with ESPaDOnS, this time must be multiplied by eight to obtain the total observation time (without overheads). 

The signal-to-noise ratio (S/N) \textit{per} CCD pixel for the order no.~35 (centered at 647~nm and near the H$\alpha$ line) is also shown in Table~\ref{tablog}. This S/N is computed by single frame using the Libre-ESpRIT pipeline and recorded in the image headers. All the objects in our sample except three have S/N higher than 150. Finally, the seeing value for each measurement is also indicated in Table~\ref{tablog}. The quoted value is the average from the values during the observation time as it appeared in the night report for each program, when available.

As mentioned before, our sample has previously published broad-band polarization and the full polarization range goes from 0.04\,\% (\object{PDS~080}) to  4.8\,\% (\object{HD~150193} and \object{HD~19820}; see Table~\ref{tabres}). We believe that this interval is appropriate for testing the continuum linear polarization in most common polarization levels of astrophysical objects.

\subsection{Circular polarization}

We selected additional objects from the CFHT archive to test the circular continuum polarization. A total of seventeen measurements from eleven objects were reduced using Opera to compute \textit{V} Stokes spectra. This subsample is also shown in Table~\ref{tablog}. The sources include two objects (HD~197770 and HD~194279) used by Cox et al. (2011) to characterize spectropolarimetric features of diffuse interstellar bands, a pole-on Be star (HD~208057, Hayes~\cite{hay80}), a non-magnetic standard star (HD~175362, Donati et al.~\cite{don97}), three linear polarized standard stars observed in \textit{V} Stokes (HD~147084,  HD~198478, HD~204827), and four bright sources to test the short integration times (Sirius, HD~62509, HD~98485, and Vega).

In general, all these sources are expected to show null circular continuum polarization. In this sense, this can help to investigate the accuracy for the \textit{V} spectra. The integration times are spanned since very short exposures (a few seconds) to long shots (up to 900~seconds). In particular, short integration times were chosen to test their possible correlation with a high instrumental polarization.

\section{Reductions with OPERA\label{reduct}}

The OPERA software\footnote{Version 2013/Oct/21.} was used to reduce the ESPaDOnS data. We installed and ran OPERA in a Ubuntu 12.03 Linux system. An overall description of the reduction steps for OPERA can be found in Martioli et al.~(\cite{mar12}). The complete source software is available from the OPERA homepage\footnote{http://sourceforge.net/projects/opera-pipeline}. 

For extracting the flux from the two polarized beams, OPERA implements the optimal extraction algorithm by Horne~(\cite{hor86}) and Marsh~(\cite{mar89}) using an oversampled tilted aperture. The aperture for extraction consists of a tilted rectangle, where the tilt angle is measured from a two-dimensional instrument illumination profile obtained from a calibration Fabry-Perot alignment exposure. An oversampled tilted aperture for flux extraction is required for ESPaDOnS, since it has an image slicer, which produces a tilted pseudo-slit as the dispersing element. In addition, ESPaDOnS has a limited CCD pixel size that undersamples the spectrum. The detailed algorithms for calibration and reduction of ESPaDOnS data implemented in the OPERA pipeline are thoroughly described in another paper in preparation (Martioli et al. 2014).

A typical calibration set for the observations presented here is composed of three bias frames, one Th-Ar lamp calibration frame, one alignment frame (Fabry Perot), and a sequence of 20 halogen lamp flat frames. For each science frame, a proper reduced calibration set must be generated considering the same observation mode (Fast, Normal, or Slow) and detector (EEV1 or Olapa).

Each Stokes parameter ($Q_{\lambda}$, $U_{\lambda}$, or $V_{\lambda}$) is obtained by reducing four single frames in proper positions of the retarders. After this process, OPERA yields the $Q_{\lambda}$, $U_{\lambda}$, or $V_{\lambda}$ spectra and the associated errors. In the case of linear polarization, these spectra were conveniently adapted to be read by the \texttt{specpol}\footnote{\texttt{specpol} was originally written by A. Carciofi and is available in PCCDPACK (Pereyra et al.~\cite{per00})} package, which allowed us to show multiplots including intensity ($I_{\lambda}$), polarization ($P_{\lambda}$), and position angle ($\rm{PA}_{\lambda}$) spectra. This package allows spectra binning using a variable bin size having a constant polarization error \textit{per} bin. The normalized intensity spectra were directly extracted  from the products yielded by the Libre-ESpRIT pipeline. Several telluric absorption bands are present in these spectra. For our purposes, the correction for telluric lines is not necessary.

\begin{figure*}
  \begin{tabular}{cc}
      \includegraphics[height=54mm,width=85mm]{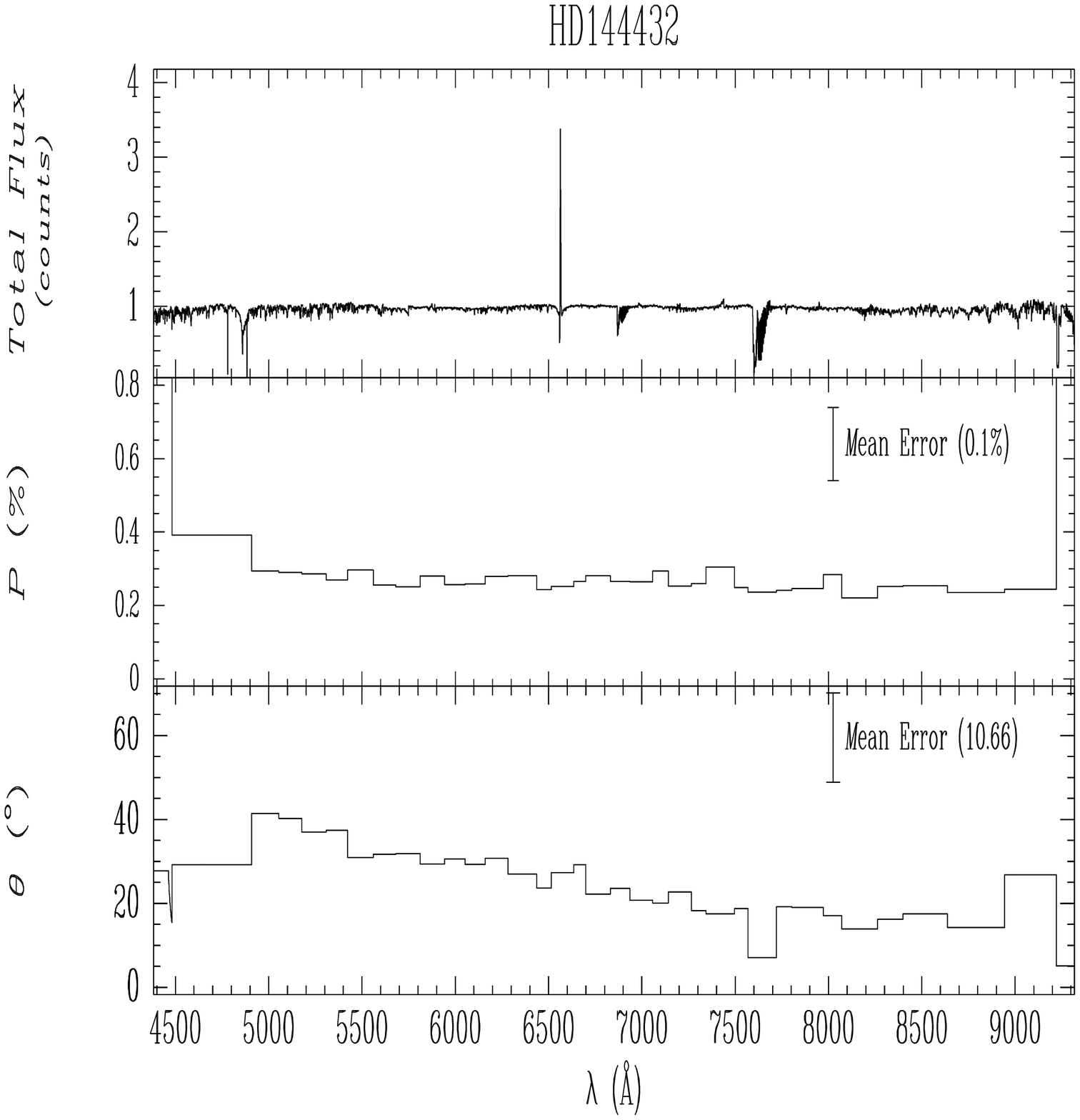}&
      \includegraphics[height=54mm,width=85mm]{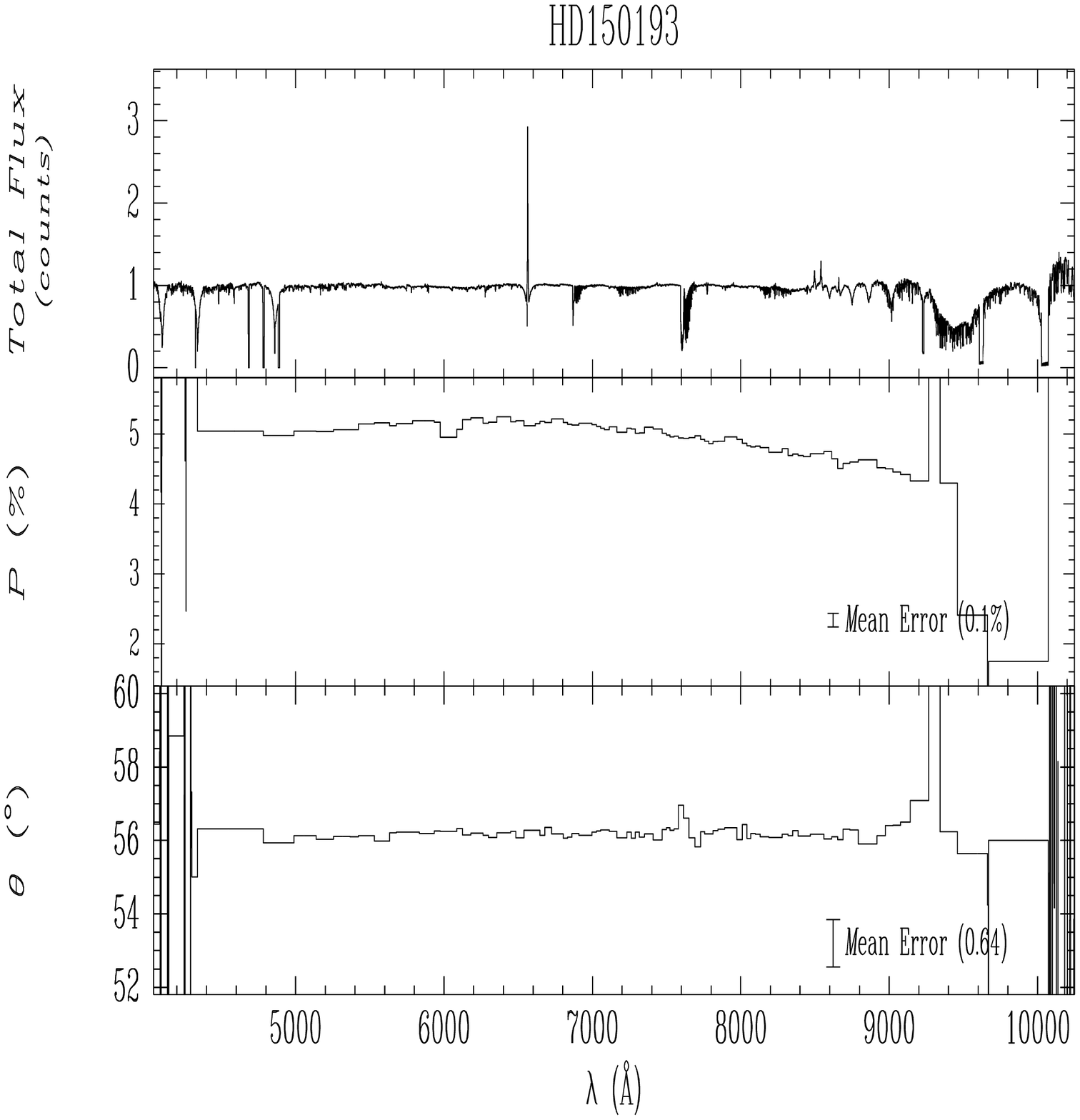}\\  
    \end{tabular}
\caption{ESPaDOnS spectropolarimetry of the HeAeBe stars \object{HD~144432} and \object{HD~150193} gathered from the CFHT archive. The spectra show the normalized total flux (top), polarization level (middle), and polarization position angle (bottom, $\theta$ = PA) binned using a variable bin size with a constant polarization error \textit{per} bin of 0.1\,$\%$.
}
\label{arcfig}
\end{figure*}

\begin{figure*}
  \begin{tabular}{cc}
    \includegraphics[height=54mm,width=85mm]{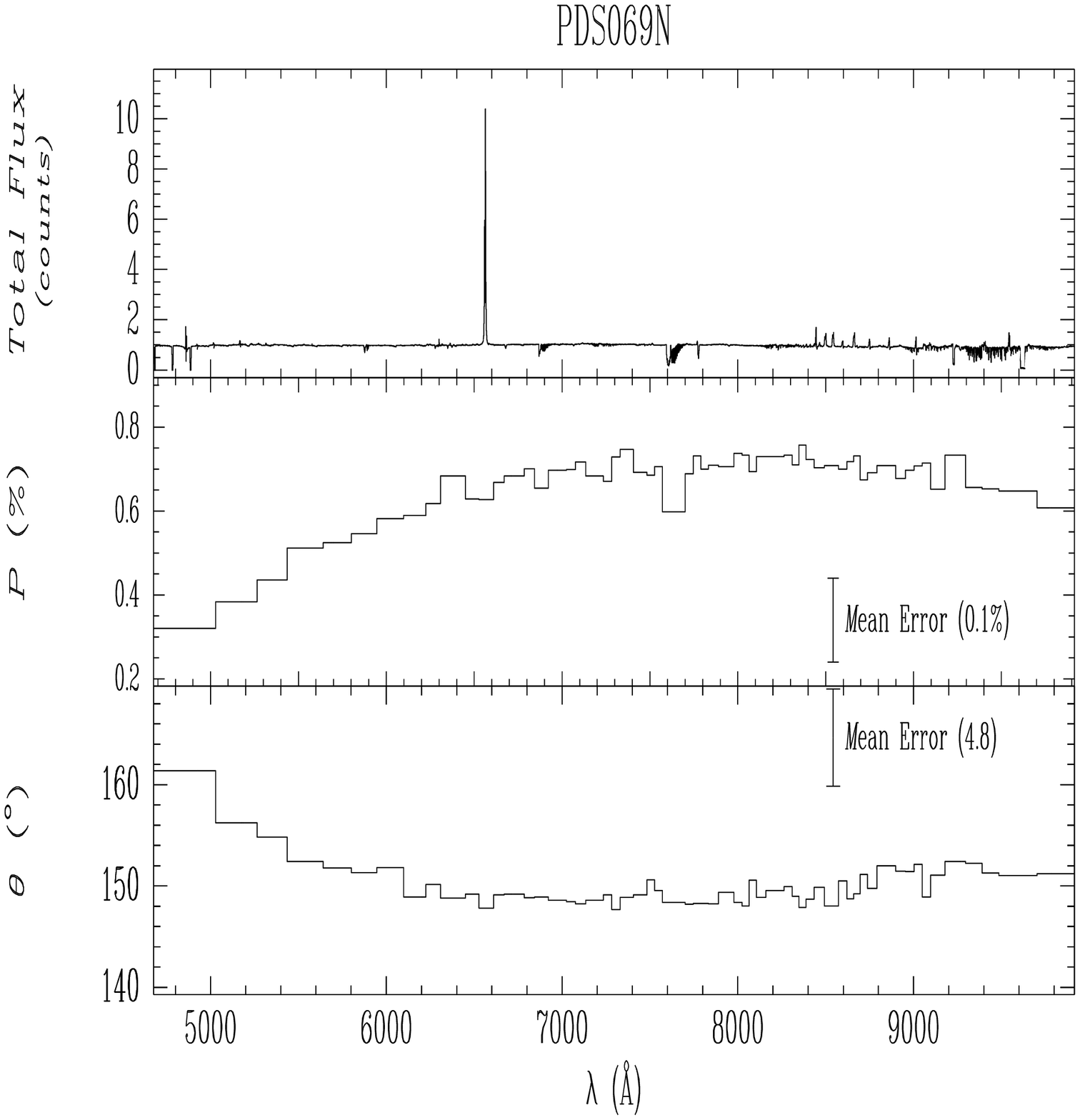}&
    \includegraphics[height=54mm,width=85mm]{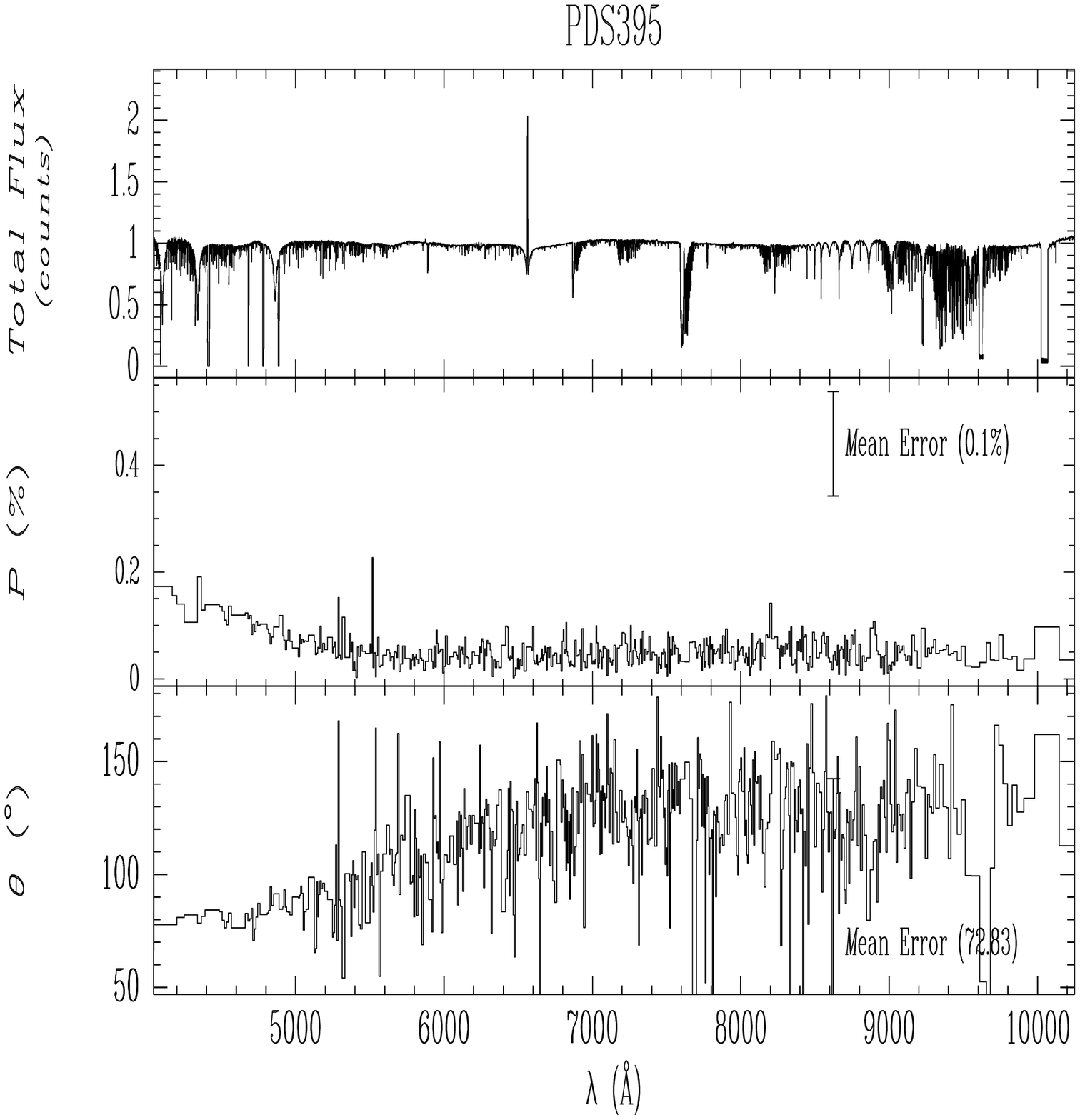}\\  
    \includegraphics[height=54mm,width=85mm]{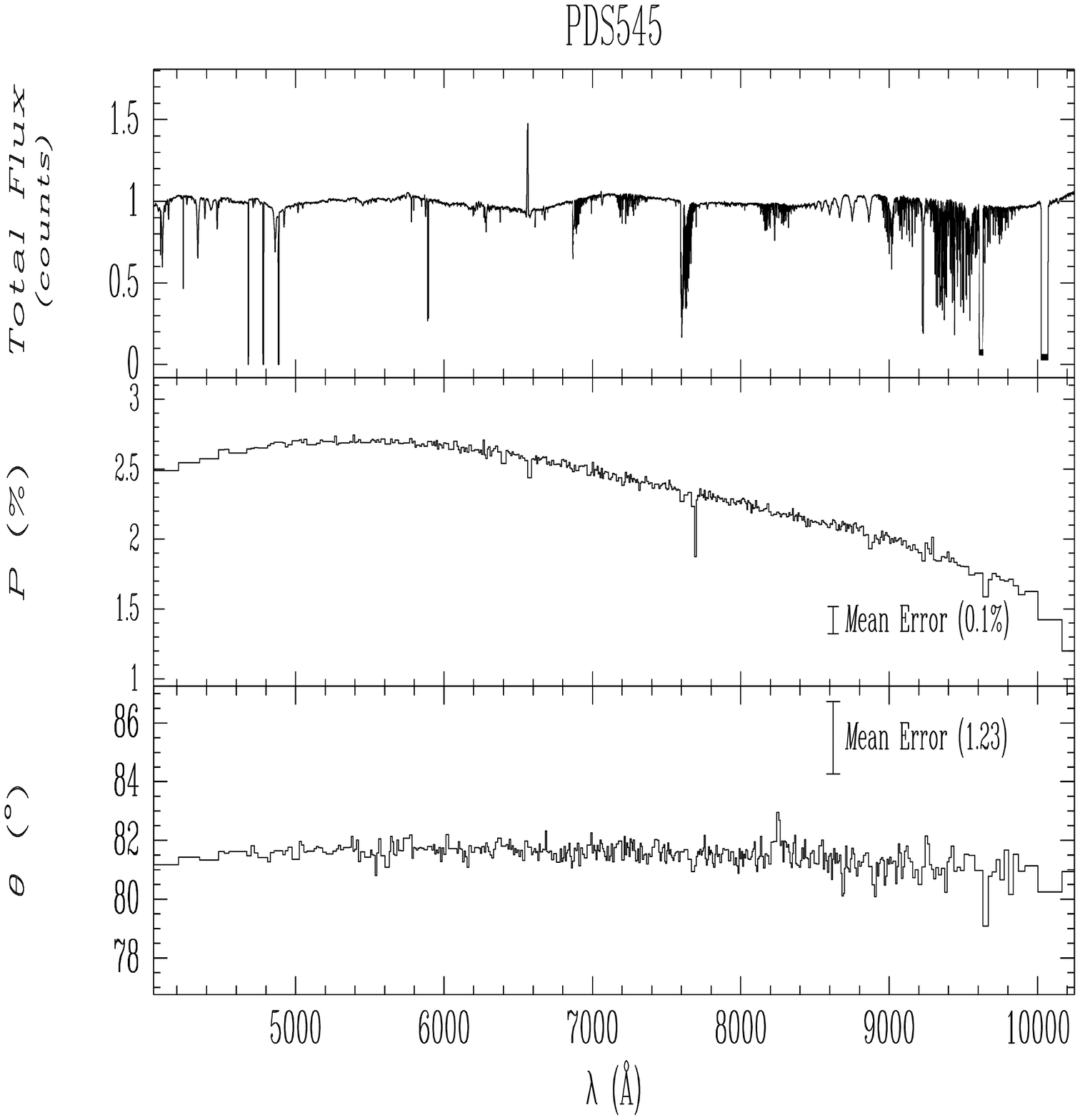}&
    \includegraphics[height=54mm,width=85mm]{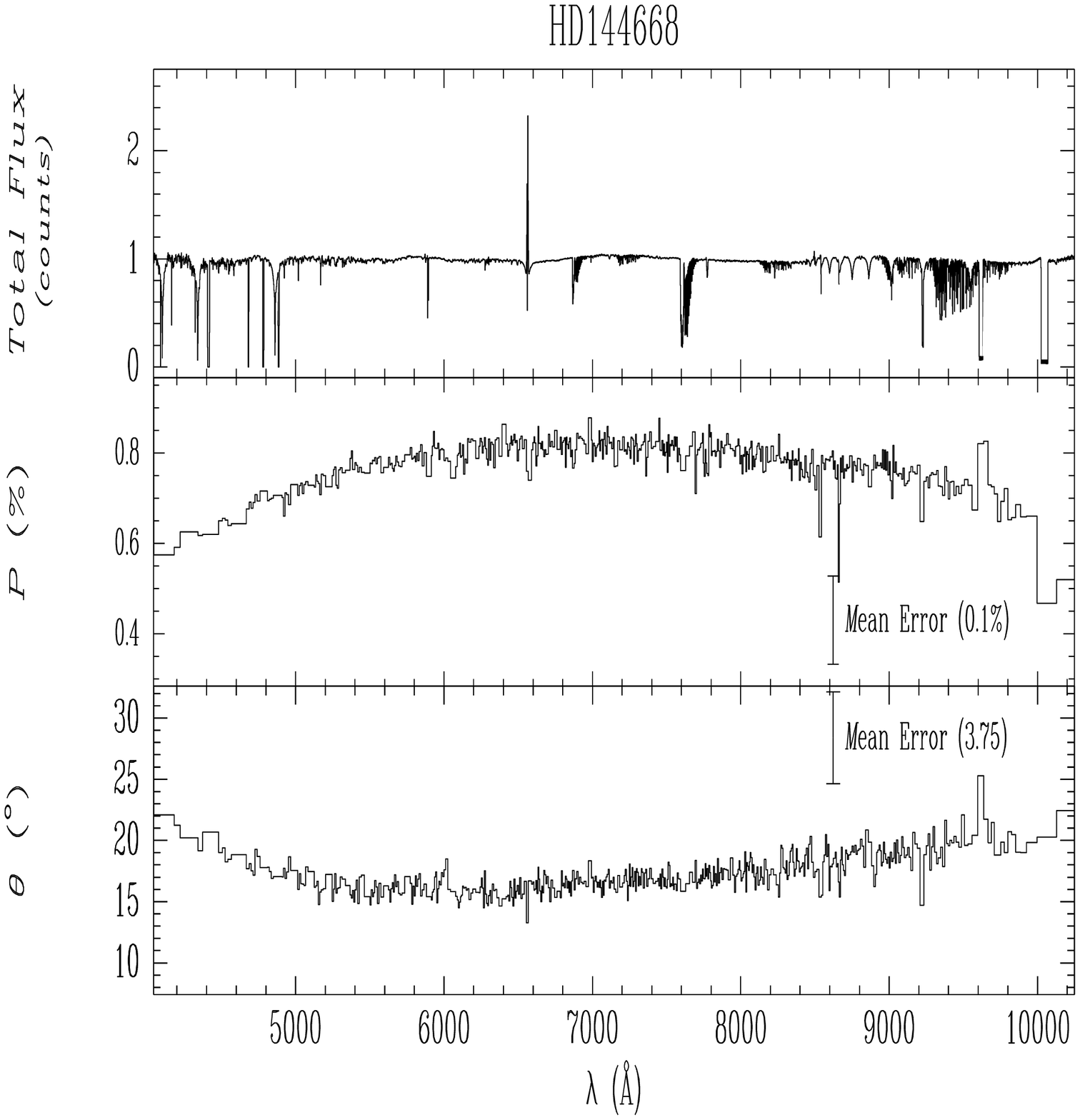}\\
    \includegraphics[height=54mm,width=85mm]{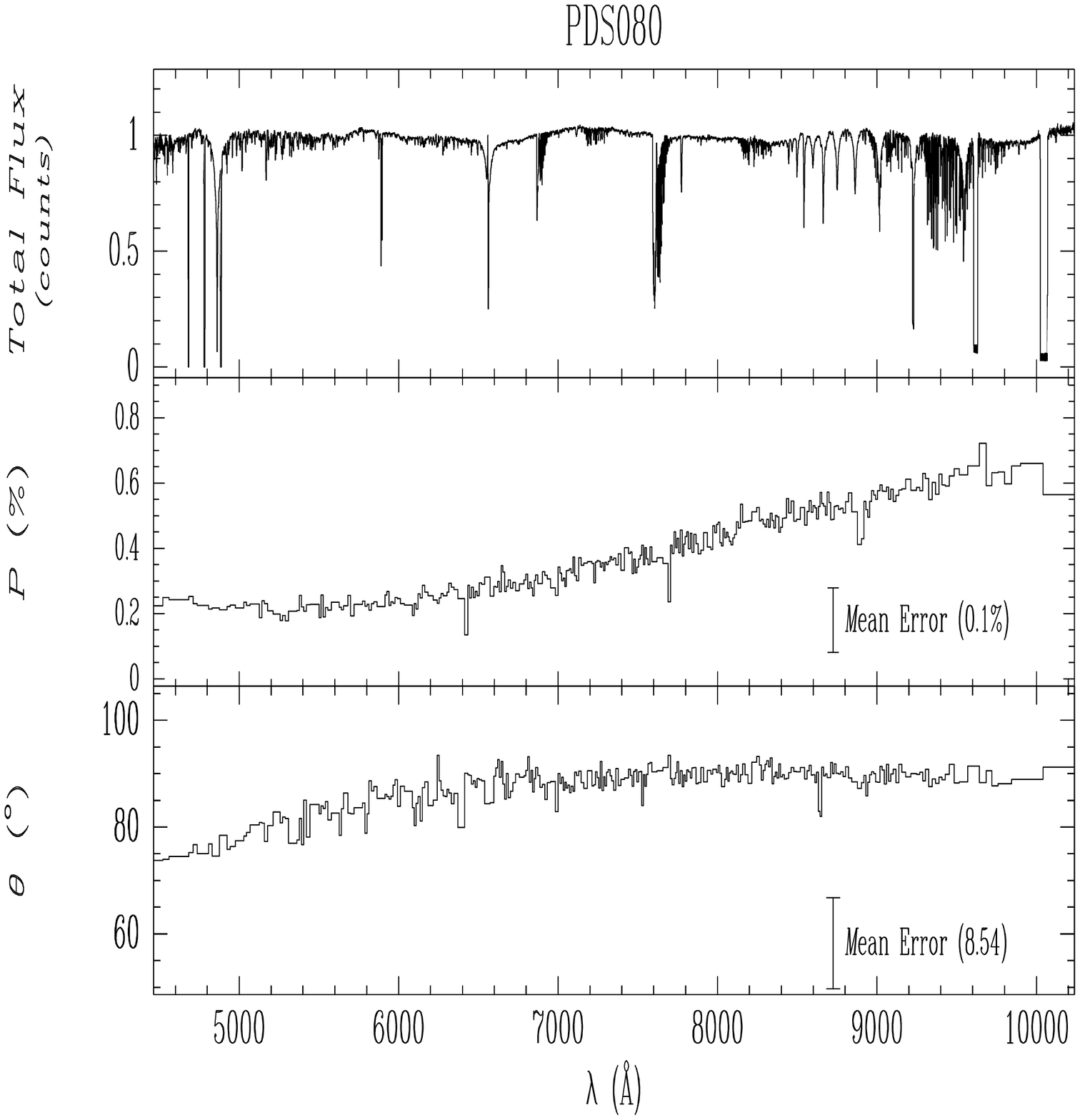}&
    \end{tabular}
\caption{As Fig.~\ref{arcfig} for the HeAeBe stars gathered with ESPaDOnS in the 11AB04 proposal.}
\label{11afig}
\end{figure*}


\begin{figure*}
  \begin{tabular}{cc}
    \includegraphics[height=54mm,width=85mm]{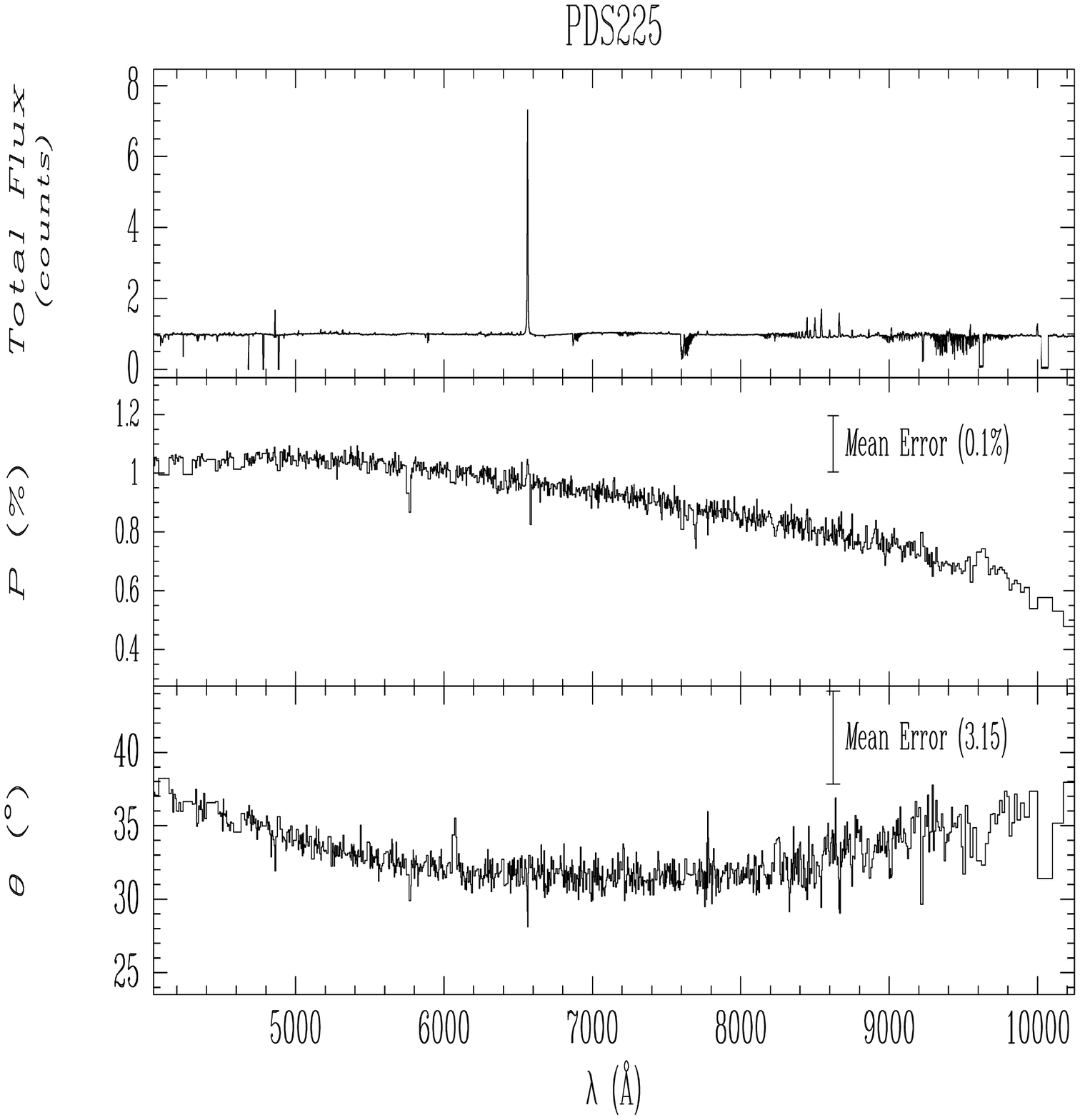}&
    \includegraphics[height=54mm,width=85mm]{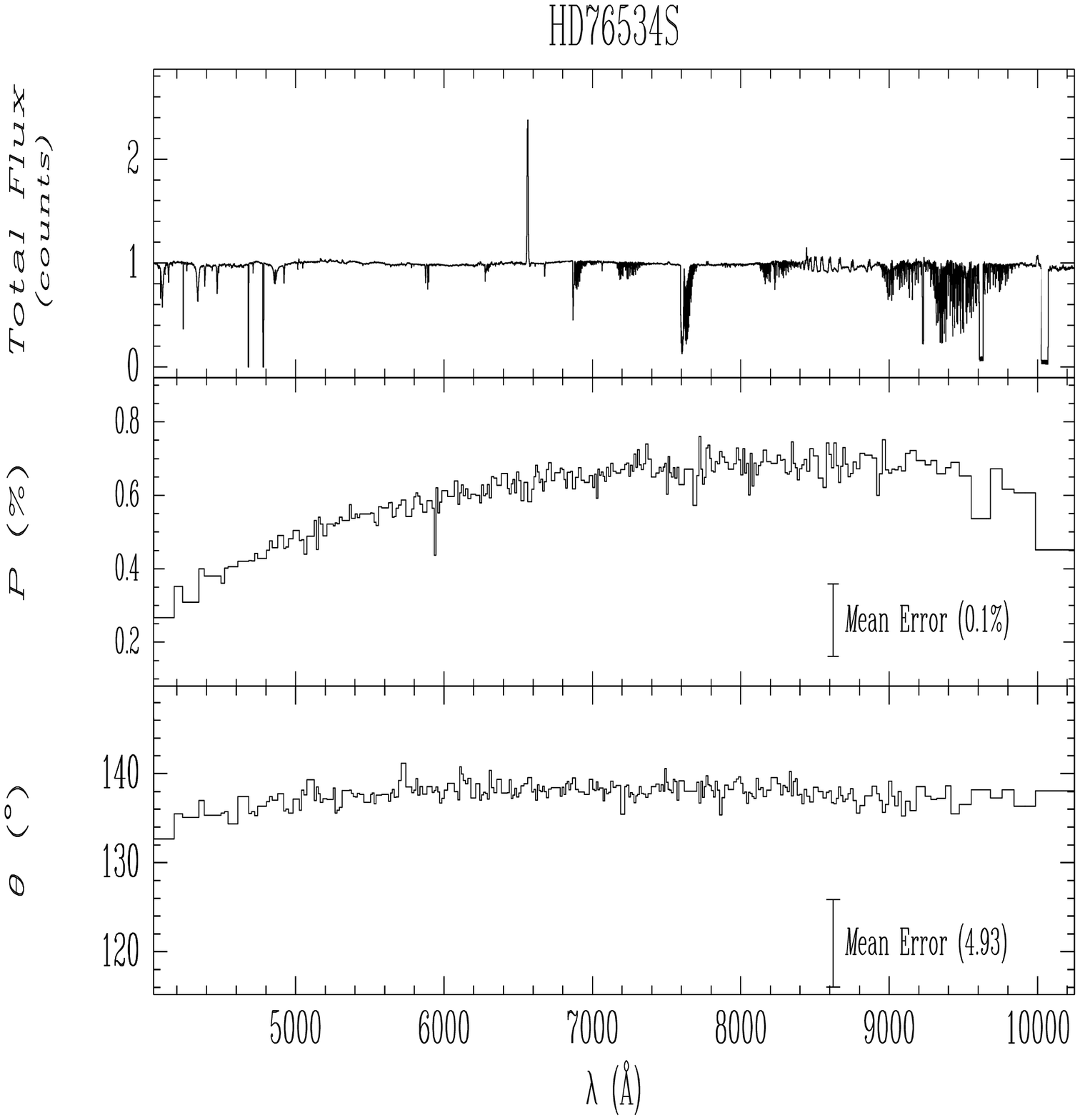}\\ 
    \includegraphics[height=54mm,width=85mm]{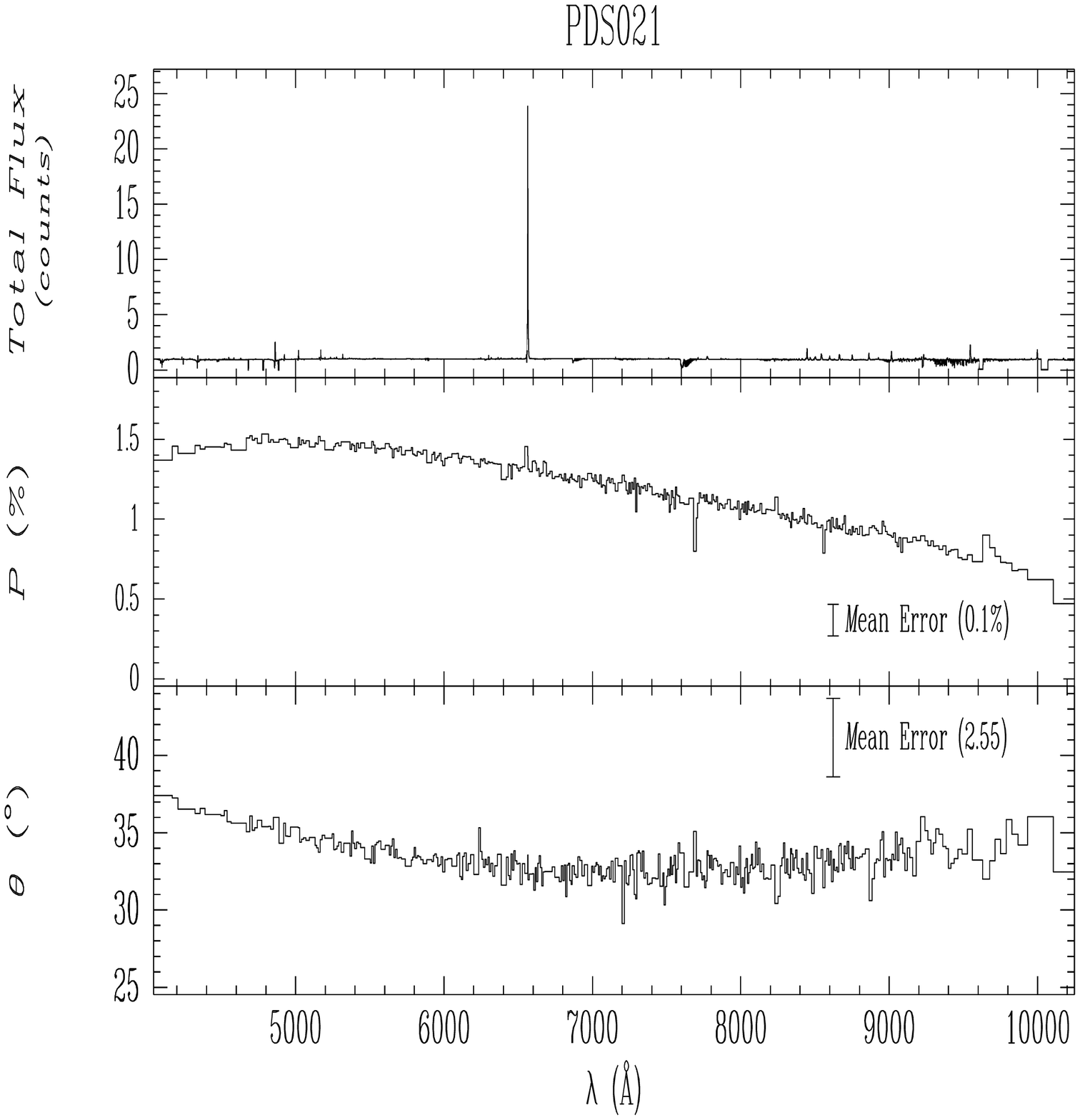}&
    \includegraphics[height=54mm,width=85mm]{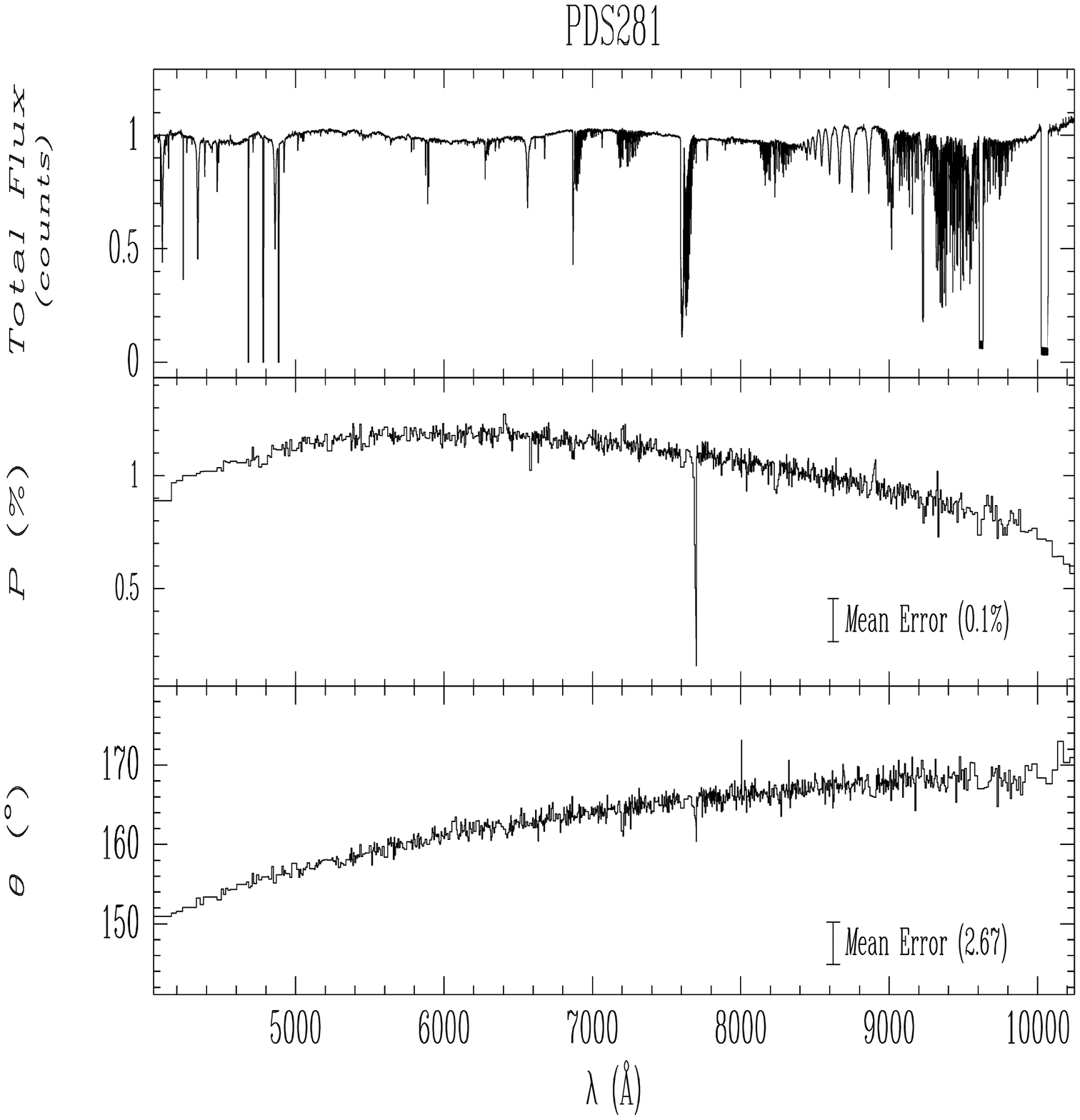}\\ 
    \end{tabular}
\caption{As Fig.~\ref{arcfig} for the HeAeBe stars gathered with ESPaDOnS in the 11BB05 proposal.}
\label{11bfig}
\end{figure*}

\begin{figure*}
  \begin{tabular}{cc}
    \includegraphics[height=54mm,width=85mm]{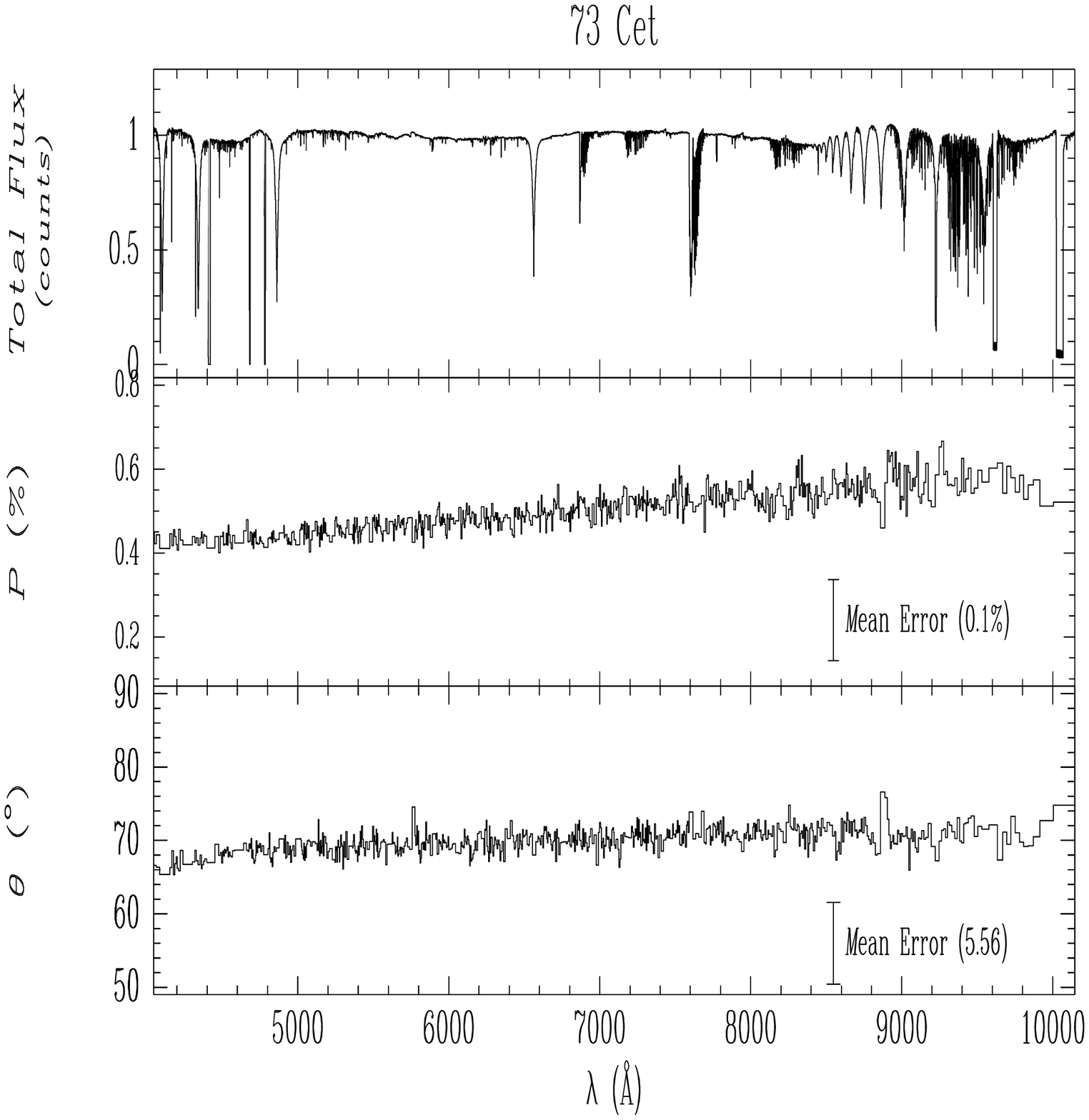}&
    \includegraphics[height=54mm,width=85mm]{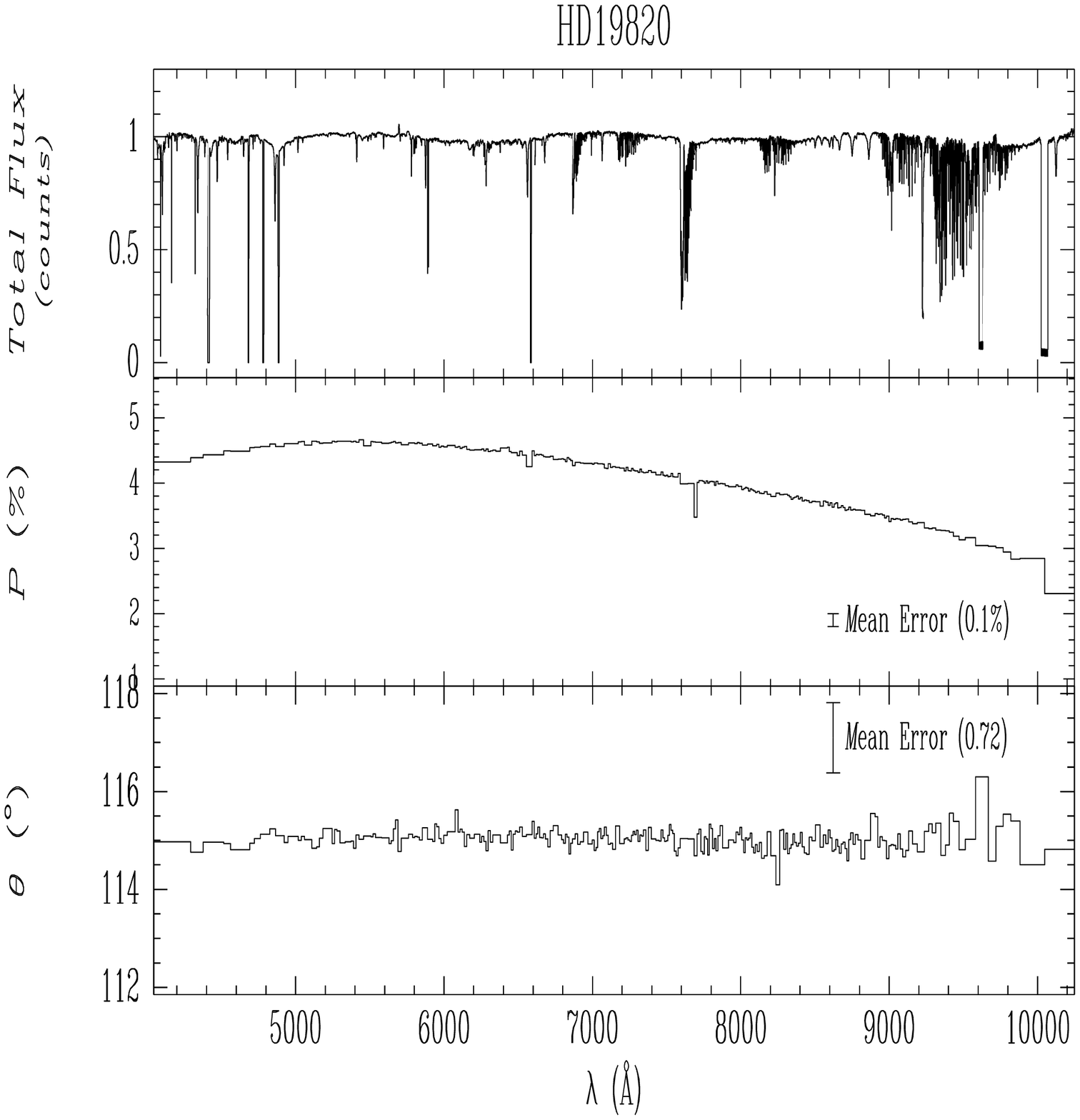}\\ 
    \end{tabular}
\caption{As Fig.~\ref{arcfig} for the standard stars \object{73~Cet} (unpolarized) and \object{HD~19820} (polarized) gathered with ESPaDOnS in the 11BE95 proposal.}
\label{padfig}
\end{figure*}

\section{Measuring the polarization}
\label{measure}
 
A detailed description of the procedure to measure the polarization using a simple polarimeter can be found in Bagnulo~et~al.~(\cite{bag09}). In summary, two different methods are available to obtain any of the Stokes parameters (\textit{Q}, \textit{U}, or \textit{V}): the \textit{ratio} and \textit{difference} methods. These two methods are implemented in the OPERA software. The linear polarization is a function of two Stokes parameters, \textit{Q} and \textit{U}, and the circular polarization is given by\textit{V}. Each of these parameters is computed using four independent exposures gathered in different positions of the retarders, which rotate the polarization plane of the incident beam. The polarimetric mode of ESPaDOnS used a combination of three Fresnel rhombs as retarders. Two of them rotate (with angles $\phi_1$ and $\phi_3$) and define the net polarization plane to be measured. In each exposure two spectra are recorded with orthogonal polarization states following the splitting done by the analyzer (a Wollaston prism).

It can be shown that the transmitted intensity of light passing through the polarimetric unit of ESPaDOnS, 
is given by

\begin{equation}
\begin{multlined}
I_\pm(\phi_1,\phi_3) = \frac{I_{*}}{2} [\;1 \pm Q \cos 4 \phi_1 \cos 4 \phi_3 \\
\pm U \sin 4 \phi_1 \cos 4 \phi_3 \mp V \sin 4 \phi_3\;],
\end{multlined}
\label{eq.itoff}
\end{equation}
\\
\noindent where $\textit{I}_{*}$ is the incident intensity, and $\phi_1$ and $\phi_3$ are the directions of the retarder's optical axes. 
The 
signals indicate the intensities of the measured orthogonal beams produced by the analyzer splitting. 
All these directions are measured with respect to the orientation of the instrument. 

In this sense, a combination of four pairs of rotations of the retarders ($\phi_1/\phi_3$ $=$ 0$\degr$/0$\degr$, 90$\degr$/45$\degr$, 45$\degr$/90$\degr$, and 135$\degr$/135$\degr$) are needed to measure $Q_\lambda$ and additional four combinations (22$\fdg$5/0$\degr$, 112$\fdg$5/45$\degr$, 67$\fdg$5/90$\degr$, and 157$\fdg$5/135$\degr$) are used to measure $U_\lambda$. For \textit{V}, four combinations are needed: (0$\degr$/67$\fdg$5, 90$\degr$/22$\fdg$5, 45$\degr$/22$\fdg$5, and 135$\degr$/67$\fdg$5). This procedure yields a set of eight spectra for each computed Stokes parameter. 

%

Following the \textit{ratio} method, we define the ratio between the orthogonal polarization intensities in a given combination of the retarders position as follows: 

\begin{displaymath}
r_{\phi_1,\phi_3} = \frac{I_+(\phi_1,\phi_3)} {I_-(\phi_1,\phi_3)}. 
\end{displaymath}
\\
Then, the Stokes parameters can be obtained from

\begin{displaymath}
Q =  \frac{ R_{Q} - 1}{ R_{Q} + 1}, \;\; \textrm{being} \;\;
R^{4}_{Q} = \left(\frac{r_{0,0}}{r_{90,45}}\right) \left(\frac{r_{135,135}}{r_{45,90}}\right),\;\;\;\;\;\;\;\;\;\;\;\;
\end{displaymath}
\begin{displaymath}
U =  \frac{ R_{U} - 1}{ R_{U} + 1}, \;\; \text{being} \;\;
R^{4}_{U} =  \left(\frac{r_{22.5,0}}{r_{112.5,45}}\right) \left(\frac{r_{157.5,135}}{r_{67.5,90}}\right), \rm{and}
\end{displaymath}
\begin{displaymath}
V =  \frac{ R_{V} - 1}{ R_{V} + 1}, \;\; \text{being} \;\;
R^{4}_{V} =  \left(\frac{r_{0,67.5}}{r_{90,22.5}}\right) \left(\frac{r_{135,67.5}}{r_{45,22.5}}\right).\;\;\;\;\;\;\;\;\;
\end{displaymath}
\\
The degree (\textit{P}) and the position angle (PA) of the linear polarization are determined from the following expressions:

\begin{equation}
P = \sqrt{Q^{2}+U^{2}} \;\;\;\; ; \;\; 
\text{PA} = \frac{1}{2} \arctan \left(\frac{U}{Q}\right).
\label{eq.ppa}
\end{equation}

In the same way, the \textit{difference} method can be applied if we define the ratio between the difference and the sum of the two orthogonal polarization intensities in a given combination of the retarders position, as follows:

\begin{displaymath}
r_{\phi_1,\phi_3} = \frac{I_+(\phi_1,\phi_3) - I_-(\phi_1,\phi_3)} {I_+(\phi_1,\phi_3) + I_-(\phi_1,\phi_3)}.
\end{displaymath}
\\
Then, the Stokes parameters can be obtained from

\begin{displaymath}
Q =  \frac{1}{4} [(r_{0,0} - r_{90,45}) - (r_{45,90} - r_{135,135})],\;\;\;\;\;\;\;\;\;\;\;\;\;\;\;\;\;
\end{displaymath}
\begin{displaymath}
\;\;\;U =  \frac{1}{4}[(r_{22.5,0} - r_{112.5,45}) - (r_{67.5,90} - r_{157.5,135})],\;\; \rm{and}
\end{displaymath}
\begin{displaymath}
\;\;\;V =  \frac{1}{4}[(r_{0,67.5} - r_{90,22.5}) - (r_{45,22.5} - r_{135,67.5})].\;\;\;\;\;\;\;\;\;\;\;\;
\end{displaymath}
\\

In this work, the \textit{ratio} method was used to compute the continuum (linear and circular) polarization using OPERA. A comparison between the two methods is presented in Sect.~\ref{compmeth}. 

\begin{figure*}[!ht]
\resizebox{18cm}{!}{\includegraphics[clip]{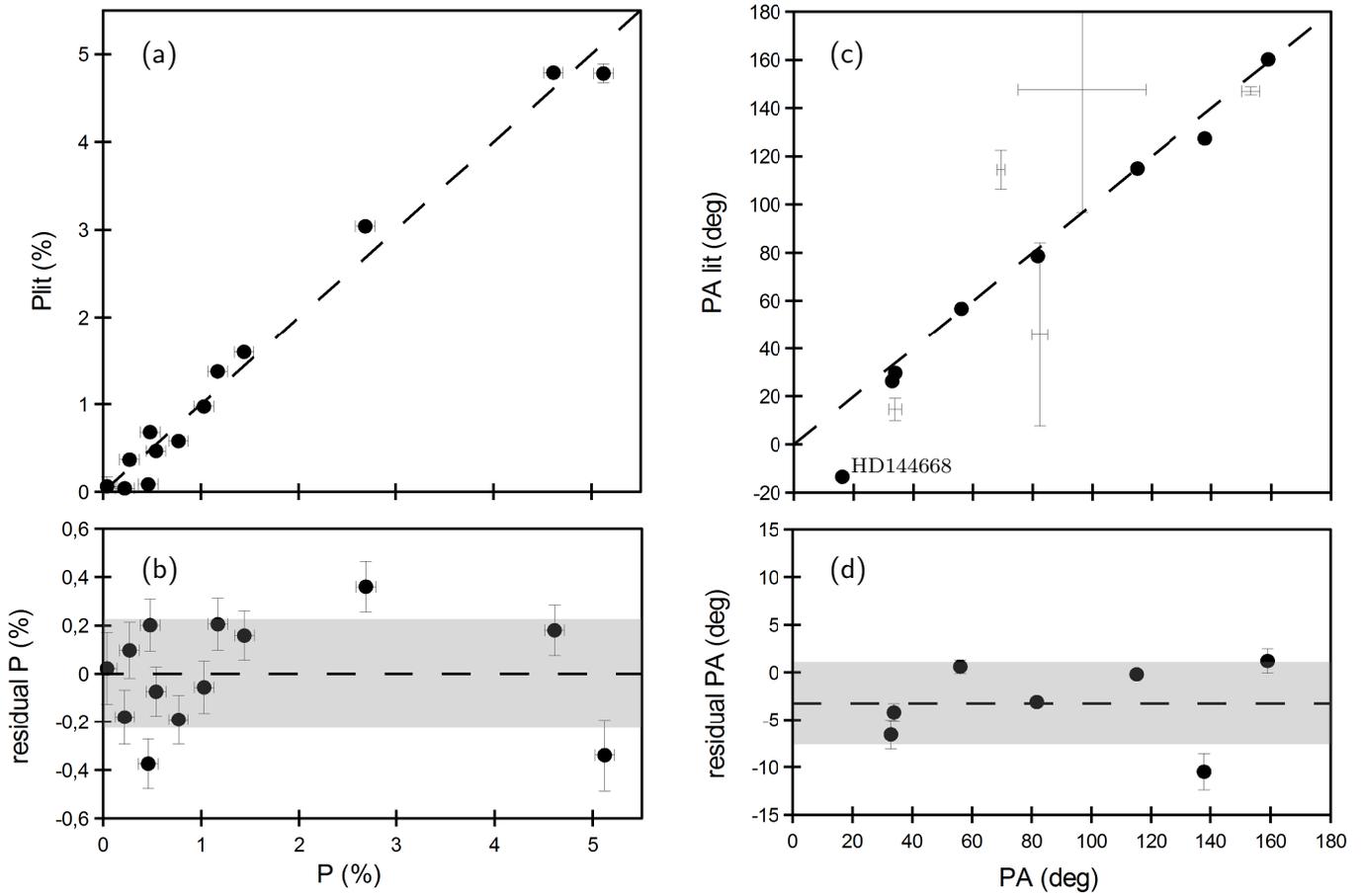}}
\caption{Comparison of synthetic \textit{V} broad-band linear polarization from ESPaDOnS and literature values. (a) The abscissa is the degree of polarization  ($\textit{P}$) measured using ESPaDOnS. The ordinate is the literature value ($\textit{P}_{\rm lit}$). The dashed line represents $\textit{P}_{\rm lit}$~=~$\textit{P}$. (b) The residual ($\textit{P}_{\rm lit}-\textit{P}$). The gray zone indicates a one-sigma dispersion around the mean value: 0~$\pm$~0.23$\,\%$ (dashed line). (c) The abscissa is the polarization position angle (PA) measured using ESPaDOnS. The ordinate is the literature value ($\rm{PA_{lit}}$). The dashed line represents $\rm{PA_{lit}}$~=~PA. The black dots are objects with $P > 0.5\,\%$ and crosses with $P < 0.5\,\%$. The PA variable \object{HD~144668} is labeled. (d) The residual ($\rm{PA_{lit}-PA}$) only for objects with $P > 0.5\,\%$. The gray zone indicates a one-sigma dispersion around the mean value: $-$3$\fdg$3~$\pm$~4$\fdg$3.
}
\label{compfig}
\end{figure*}

\begin{figure*}[!ht]
\resizebox{18cm}{!}{\includegraphics[clip]{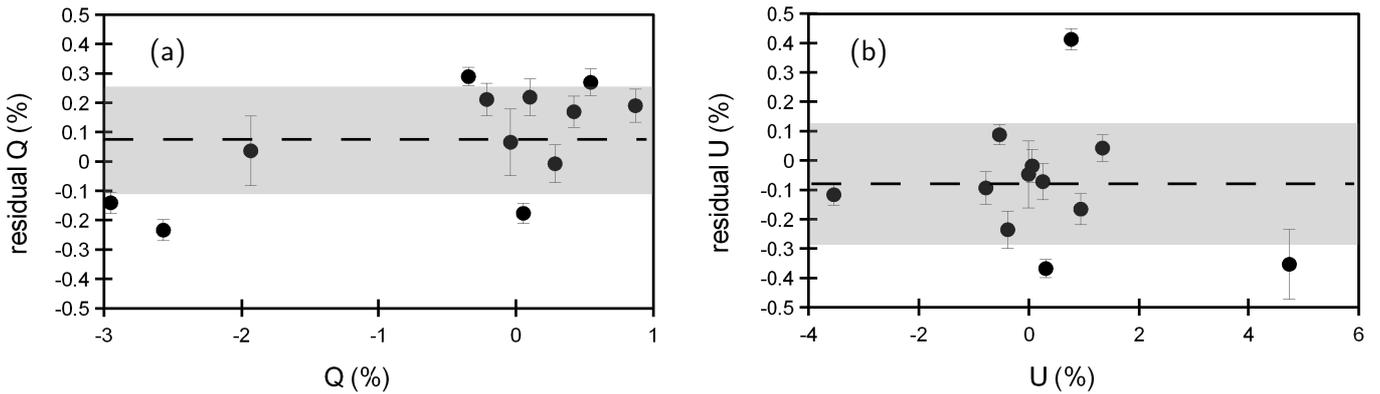}}
\caption{Residuals in Stokes parameters between synthetic \textit{V} broad-band linear polarization from ESPaDOnS and literature values. (a) The abscissa is the \textit{Q} Stokes parameter measured using ESPaDOnS. The ordinate is the residual ($\textit{Q}_{\rm lit}-\textit{Q}$). The gray zone indicates a one-sigma dispersion around the mean value: 0.07~$\pm$~0.18$\,\%$ (dashed line). (b) The abscissa is the \textit{U} Stokes parameter measured using ESPaDOnS. The ordinate is the residual ($\textit{U}_{\rm lit}-\textit{U}$). The gray zone indicates a one-sigma dispersion around the mean value: -0.08~$\pm$~0.21$\,\%$ (dashed line). }
\label{stokesfig}
\end{figure*}

\begin{figure*}[ht!]
\resizebox{18cm}{!}{\includegraphics[clip]{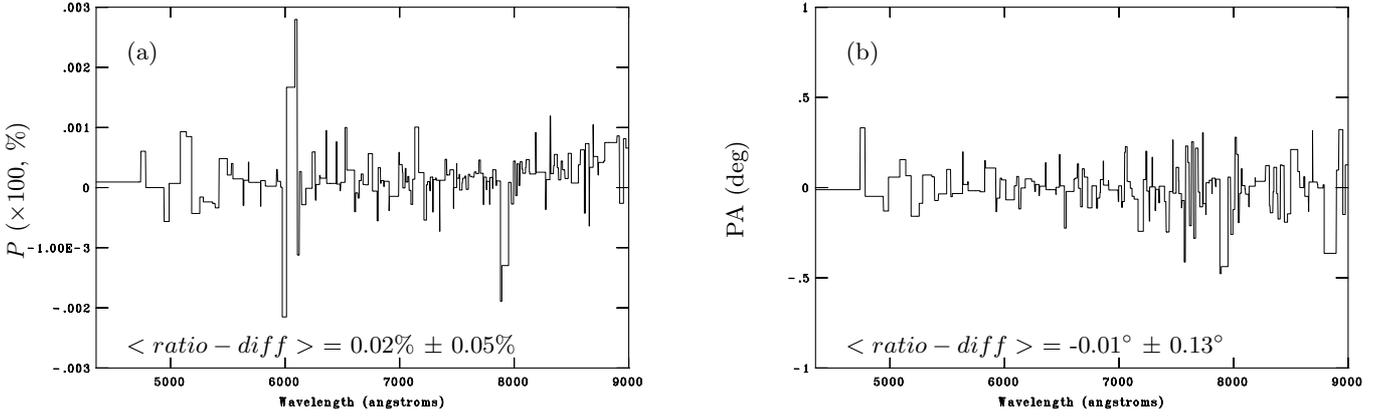}}     
\caption{Residuals between the ratio and difference methods computed by OPERA for the high polarized object \object{HD~150193}. (a) degree of polarization; (b) position angle. The spectra are binned using a variable bin size with a constant polarization error \textit{per} bin of 0.1$\,\%$. The mean values and one-sigma dispersion are indicated in each case.
}
\label{compmets}
\end{figure*}

\begin{figure*}[!ht]
\resizebox{18cm}{!}{\includegraphics[clip]{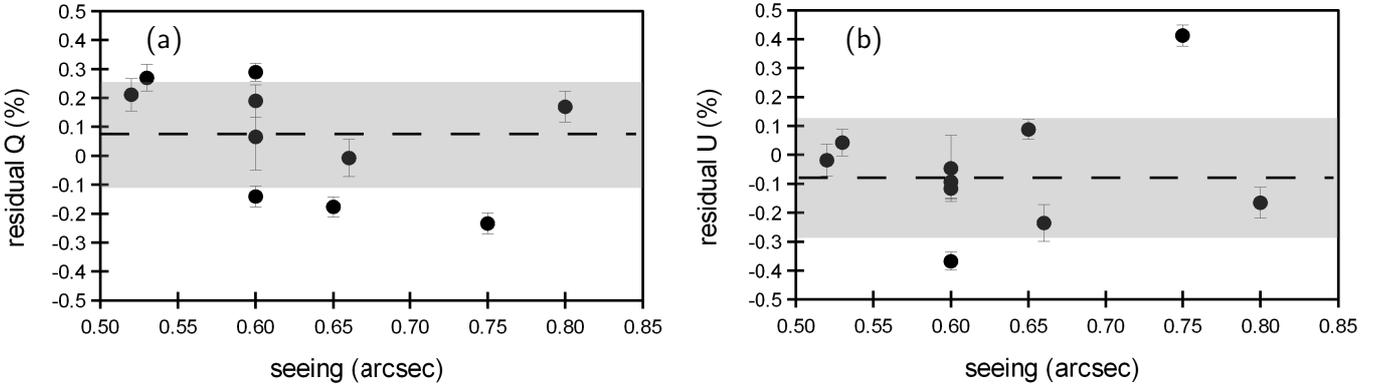}}
\caption{Residuals in Stokes parameters versus seeing values for \textit{Q} (a) and \textit{U} (b). The dashed line and gray zones are as in Fig.~\ref{stokesfig}.}
\label{seeingfig}
\end{figure*}

\begin{figure*}[!ht]
\resizebox{18cm}{!}{\includegraphics[clip]{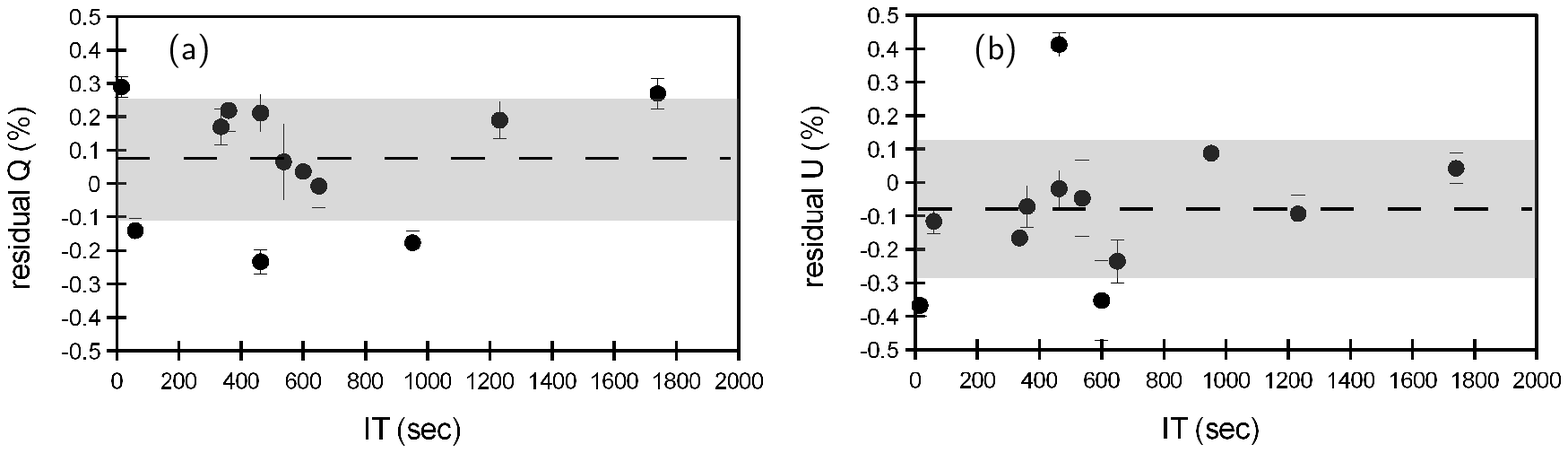}}
\caption{Residuals in Stokes parameters versus integration times for \textit{Q} (a) and \textit{U} (b). The dashed line and gray zones are as in Fig.~\ref{stokesfig}.}
\label{itimefig}
\end{figure*}

\section{Results\label{results}}

\subsection{Linear polarization}

Figures~\ref{arcfig}-\ref{padfig} show the results for the spectropolarimetry of our linear polarization sample (Table~\ref{tablog}). The spectra are binned to enhance the signal-to-noise ratio. We selected a polarimetric error of 0.1\,\% \textit{per} bin as a good compromise to show the data properly. 

Our sample includes HeAeBe stars with very low polarization levels such as \object{HD~144432}, \object{PDS~395}, and \object{PDS~080}. Interestingly, the mean level of the binned polarization spectrum of \object{PDS~395} (Fig.~\ref{11afig}) is lower than 0.1\,\%, which is a good example of the ability of ESPaDOnS to detect low polarizations. On the other hand, high polarization levels (up to $\sim$5\,\%) are well sampled, as are the cases of the HeAeBe star \object{HD~150193} (Fig.~\ref{arcfig}) and the polarized standard star \object{HD~19820} (Fig.~\ref{padfig}).

HeAeBe stars are known to exhibit time-variability in flux and polarization. 
Hence caution must be taken when comparing measurements, because possible differences may be intrinsic to the object and not due to observational uncertainties. We looked for previous broad-band polarization measurements of our HAeBe sample to check for variability and only in one case confirmed variability was found (see below).

In order to facilitate the comparison with the literature and test the quality of the continuum linear polarization from ESPaDOnS, we calculate synthetic \textit{VRI} broad-band polarization averaging the Stokes parameters ($\textit{Q}_\lambda$ and $\textit{U}_\lambda$) between the cut-on and cut-off wavelengths associated with these standard broad-band filters. The intervals were 5070$-$5950$\rm{\,\AA}$, 5890$-$7270$\rm{\,\AA}$, and 7315$-$8805$\rm{\,\AA}$ for the \textit{V}, \textit{R}, and  \textit{I} filters, respectively. The synthetic values are shown in Table~\ref{tabres} along with the literature values when available. 

All the HeAeBe stars in our sample have optical polarization (at \textit{V} filter) previously measured by Rodrigues et al.~(\cite{rod09}). The two polarimetric standard stars (\object{73~Cet} and \object{HD~19820}) have published values by Schmidt et al.~(\cite{sch92}). 

In Figure~\ref{compfig}, we compare the polarization level at \textit{V} band of the objects in our sample with their published values ($P_{lit}$) as given in Table~\ref{tabres}. The polarization degree comparison (Fig.~\ref{compfig}a) shows a good agreement between ESPaDOnS data and the results published in the literature. The polarimetric accuracy is preserved for lower and higher values of polarization. The slope one is included to help the comparison. Considering all objects in the sample, the residual of the polarization measurements (Fig.~\ref{compfig}b) has a mean zero with a one-sigma dispersion of 0.23\,\%. In a few cases (3 out of 13), residuals of 0.4\,\% are found. This result gives us confidence that accuracies as low as 0.2$-$0.3\,\% can be obtained measuring the continuum linear polarization with ESPaDOnS. 

As discussed in Sect.~\ref{instrum}, small ITs must prevent a reliable measurement of the continuum polarization. The polarization degree measured by ESPaDOnS for the unpolarized standard star \object{73~Cet} is around 0.5\,\% along the full spectrum (Fig.~\ref{padfig}, \textit{left}). 
This level is well above the published polarization (0.1\,\%, Schmidt et al.~\cite{sch92}) considering our quoted errors. The S/N for this object (see Table~\ref{tablog}) is similar to other well sampled objects in our sample and this does not seem to be the origin of the discrepancy. However, an inspection of the integration times reveals that this object is the only one in our sample with a small IT (15~sec). Similar ITs were used by the ESPaDOnS team during the earlier commissioning\footnote{\tiny{http://www.cfht.hawaii.edu/Instruments/Spectroscopy/Espadons/ ContiPolar/}}. 
We believe that such short ITs are not sufficient to average out the modal noise present in the optical fibers. This is confirmed by the results for circular continuum polarization presented in the next section.


We also computed the residuals between the ESPaDOnS polarization position angle and the published values (Fig.~\ref{compfig}c). Caution must be taken with this comparison because low polarization levels cause high indetermination in the PA. For this reason, we computed the residuals in PA only for objects with a high degree of polarization (larger than 0.5\,\%, Fig.~\ref{compfig}d). Using this criterion, five objects were removed from the subsequent analysis.

In particular, the PA residual of \object{HD~144668} ($\sim$30$\degr$) also fitted our criteria of higher $P$. Nevertheless, this object is the only source in our sample that seems to show historical intrinsic variability in PA\footnote{See legends of Table~\ref{tabres}.} (Hutchinson~et~al.~\cite{hut94}, Bhatt~\cite{bha96}) and it was not included in Fig.~\ref{compfig}d. 

The residuals in PA have a mean of $-$3$\fdg$3~$\pm$~4$\fdg$3. Therefore, the accuracy in PA for measurements of the continuum linear polarization using ESPaDOnS is better than 5$\degr$. In particular, the constancy of the PA along the spectra for the polarized standard star (\object{HD~19820}, Fig.~\ref{padfig}, \textit{right}) and its comparison with the published broad-band values (see Table~\ref{tabres}) indicate a very reliable result. 

In order to confirm this issue, we computed the residuals between the \textit{V} broad-band polarization measured by ESPaDOnS and the literature values using the Stokes parameters (\textit{Q} = \textit{P}$\times$cos(2$\rm{PA}$), \textit{U} = \textit{P}$\times$sin(2$\rm{PA}$)). Figure~\ref{stokesfig}a shows the residual in \textit{Q} with a mean of 0.07$\%~\pm$~0.18$\%$. The residual in \textit{U} (Fig.~ \ref{stokesfig}b) has a similar dispersion with a mean of -0.08$\%~\pm$~0.21$\%$. In this analysis, the star HD~144668 has also been excluded for the reasons explained above. These results confirm our previous results, where accuracies associated with the polarization are 0.2$-$0.3\%. The mean values of \textit{Q} and \textit{U} are consistent with zero, showing no bias to positive or negative values, within the dispersion. These mean values can be interpreted as maximum limits for the instrumental polarization in \textit{Q} and \textit{U} parameters.

In general, the comparison of the optical synthetic broad-band polarimetry indicates that accurate continuum linear polarization measurements are feasible using ESPaDOnS. The measured polarizations are consistent with the literature values. 

\subsubsection{Comparing \textit{ratio} and \textit{difference} methods\label{compmeth}}

A comparison between the two methods implemented by OPERA (see Sect.~\ref{measure}) to compute the polarization was done for the highly polarized object \object{HD~150193}. Figure~\ref{compmets} shows the residuals between the ratio and difference methods computed in polarization degree and position angle. The residual in polarization degree is 0.02\,$\%~\pm~$0.05\,$\%$ and in polarization angle is -0$\fdg$01~$\pm$~0$\fdg$13. These values imply that both methods yield similar results, whose differences are below 0.05\,\% in polarization and $0\fdg2$ in PA. Hence, the results using OPERA are independent of the method used to compute the polarization. This is consistent with the analytical approach from Bagnulo~et~al.~(\cite{bag09}).

\subsubsection{Polarization gradients}

HeAeBe stars with unknown continuum polarization are not appropriate to study polarization gradients because they can have intrinsic polarization which is usually variable with wavelength. Hence we based our discussion on this topic using HeAeBe stars with small values of polarization, consistent with null polarization.

Figures~\ref{arcfig}$-$\ref{padfig} reveal non-negligible gradients in the polarization levels. This is especially important for low polarization levels. Considering the accuracies computed using the \textit{V} broad-band and extrapolating them to lower and higher wavelengths, a conservative polarization level of 0.6\,\% can be used as a threshold for a real detection of a particular wavelength dependence in polarization. This level must be taken with caution considering the signal-to-noise ratios used in our sample. Consistent with this, a dependence of the polarization position angle with wavelength is evident for objects with low polarizations at the level of the ESPaDOnS accuracy ($\sim$0.3\,\%). Therefore, the gradients observed in the position angle for HD~144432, PDS~395, and PDS~080 are examples of artifacts.

Interestingly, the wavelength ($\lambda_{\rm max}$) for maximum polarization ($P_{\rm max}$) for the polarized standard star in our sample (\object{HD~19820}) computed from the ESPaDOnS spectra ($\lambda_{\rm max}$~=~5298\,$\rm \AA$, $P_{\rm max}$~=~4.63$\,\%$) is in excellent agreement with the published values by Schmidt et al.~(\cite{sch92}, $\lambda_{\rm max}$~=~5235\,$\rm \AA$, $P_{\rm max}$~=~4.76\,$\%$). In this case, the consistency in $P_{\rm max}$ is better than 0.15\,\%. Moreover, the PA of HD~19820 is remarkably constant along the whole spectrum, as expected for a polarized standard star. This indicates that a PA gradient, if present, should be small. This shows that reliable wavelength dependence in polarization can be obtained for polarized objects measured by ESPaDOnS.

\subsubsection{Seeing and integration time dependence\label{seeingit}}

As discussed in Sect.~\ref{instrum}, the seeing can affect the quality of a polarization measurement. In order to verify any trend of this kind, we check the dependence of the residuals of our measurements in each Stokes parameters with the seeing value quoted in Table~\ref{tablog}. This is shown in Fig.~\ref{seeingfig}. The interval of seeing sampled in our measurements, between 0$\farcs$5~and 0$\farcs$8, is very typical of CFHT conditions. No correlation is apparent in \textit{Q} or \textit{U} Stokes parameters. This result suggests that the accuracy for the linear continuum polarization does not vary considerably with the seeing, at least for the interval covered by our sample.

Finally, the correlation between integration times and instrumental polarization in \textit{Q} and \textit{U} spectra was tested with the ITs quoted in Table~\ref{tablog}. Figure~\ref{itimefig} shows a plot of the residuals in \textit{Q} and \textit{U} versus IT. The minimum IT used in our sample was 15\,sec for the unpolarized standard star 73 Cet. This object was discussed above considering its higher residual in the polarization level. Interestingly, this is one of the two objects with both residuals (in \textit{Q} and \textit{U}) being above one-sigma outside the mean values. The second object is the polarized HeAeBe star PDS 545 which was observed with $\sim$460\,sec. Considering that small variations in polarization are common in HeAeBe stars, the measurement of 73 Cet seems to indicate a real higher instrumental polarization associated with low integration times. Nevertheless, additional data are needed to confirm this question (see next section).  

\begin{table}[!ht]
\caption{Synthetic polarimetry for \textit{V} Stokes parameter using ESPaDOnS data.}            
\label{tabvres}                          
\begin{tabular}{l  c  c}        
\hline\hline
\\[-1ex]                
Object             & IT       & \textit{V}      \\
                   & (sec)    & (\%)            \\
\hline\\[-1ex]
\object{HD~197770}          &  760  & $-$0.26 (0.03)    \\ 
\object{HD~194279}          &  730  & $+$0.29 (0.03)    \\
                            &  730  & $+$0.11 (0.02)    \\
\object{HD~208057}          &  200  & $+$0.02 (0.08)    \\
                            &  200  & $-$0.06 (0.10)    \\
\object{HD~175362}          &  120  & $-$0.15 (0.11)    \\
\object{HD~147084}          &   20  & $-$0.28 (0.08)    \\
\object{HD~198478}          &   60  & $+$0.27 (0.03)    \\
\object{HD~204827}          &  900  & $-$0.07 (0.02)    \\
\object{HD~48915 (Sirius)}  &    1  & $+$0.51 (0.06)    \\    
                            &  0.8  & $+$0.18 (0.05)    \\    
                            &  0.6  & $-$0.76 (0.06)    \\    
\object{HD~62509}           &    6  & $+$0.21 (0.08)    \\
\object{HD~89485}           &   12  & $+$0.42 (0.25)    \\
\object{HD~172167 (Vega)}   &    1  & $-$3.62 (0.24)    \\
                            &    2  & $+$0.27 (0.05)    \\     
                            &    4  & $+$0.04 (0.03)    \\ 
\\[-1ex]
\hline\\[-1ex]
\end{tabular}
\\
Errors in parenthesis.\\
\end{table}

\subsection{Circular polarization}

\begin{figure}[!ht]
\includegraphics[width=8.5cm,clip]{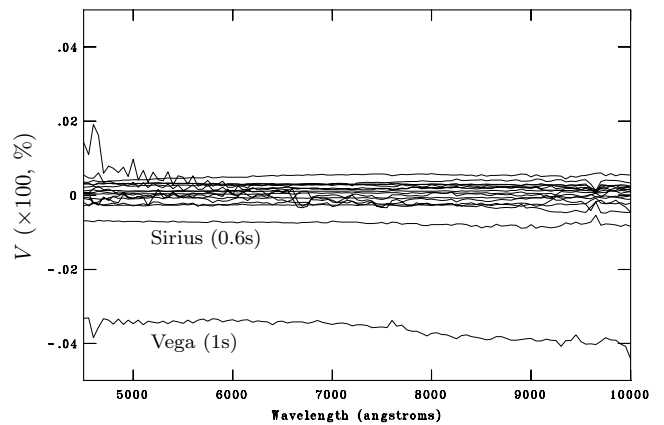}
\caption{Circular continuum polarization spectra of selected objects from the CFHT archive. Seventeen measurements are shown. The highest levels of \textit{V} in two short exposures are indicated with the integration times in parenthesis.}
\label{vcomp}
\end{figure}

Figure~\ref{vcomp} shows the circular polarization spectra for the seventeen measurements from Table~\ref{tablog}. As mentioned before, our sample includes measurements using short and long exposures. Interestingly, the short exposures show the highest levels for the \textit{V} Stokes spectra. One of the bright sources (Vega) has three different measurements with IT of 1, 2, and 4\,sec, respectively. Figure~\ref{vcomp} clearly shows that the 1\,sec measurement has the highest \textit{V} level. The same effect is also observed in another bright source (Sirius) with three measurements with different ITs. The short exposure (0.6\,sec) also shows an extreme \textit{V} value. These particular spectra are indicated in Figure~\ref{vcomp}.

\begin{figure}[!ht]
\includegraphics[width=8.5cm,clip]{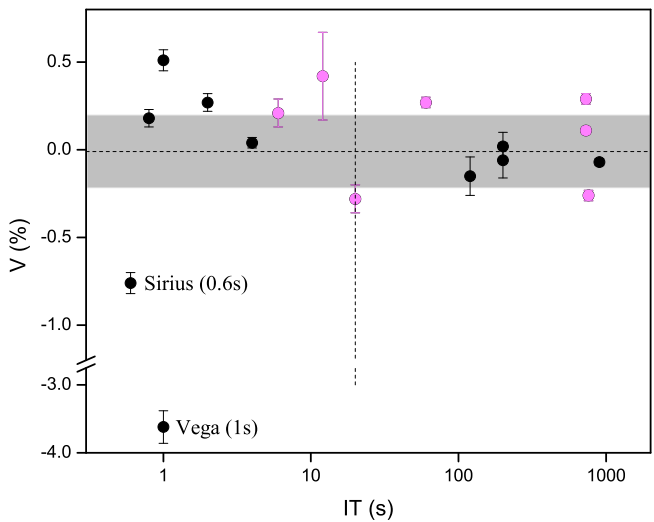}
\caption{Synthetic circular polarization for selected objects from the CFHT archive. As in Fig.~\ref{vcomp}, the highest levels are shown. The gray zone indicates a one-sigma dispersion around the mean value: -0.01 $\pm$ 0.21\,$\%$, for objects with IT $\geq$ 20\,sec. This limit is indicated by a vertical dashed line. The objects measured using the old ADC (in 2006$-$2008 interval) are shown with magenta dots. The more recent measurements are shown with black dots.}
\label{vxitime}
\end{figure}

In order to quantify the instrumental polarization in circular polarization for the ESPaDOnS data, we computed synthetic polarization averaging the \textit{V} Stokes parameter considering  the full wavelength interval available in ESPaDOnS (between 4500$-$10000\,$\rm{\AA}$). The results are shown in Table~\ref{tabvres}. Figure~\ref{vxitime} shows the circular polarization level versus the integration time. All the measurements with integration times higher than 20\,sec are below 0.3\,\% (in absolute value). A higher dispersion happens for short IT. For IT $\geq$ 20\,sec, the average is practically zero (-0.01\,\%) with a dispersion of 0.2\,\%. This is indicated by the dotted line and gray zone in Fig.~\ref{vxitime}. We confirmed that ESPaDOnS has an instrumental polarization for circular continuum polarization between 0.2$-$0.3\,\%. This level is similar to our results in linear continuum polarization. Finally, evidence of a higher instrumental polarization for short integration times is also found.

As discussed in Sect.~\ref{instrum}, a possible source of instrumental polarization was the old ADCs. To test this hypothesis, the data obtained using the ADC installed in ESPaDOnS from 2006 June to 2008 Oct are highlighted in Fig.~\ref{vxitime}. These data actually show a higher dispersion than data obtained with the present ADC, indicating that ADC can be indeed the cause of spurious polarization in old ESPaDOnS data.
\\
\\

\section{Perspectives}
\label{perps}

Assuming that the continuum polarimetry is feasible using ESPaDOnS, the range of scientific applications of this instrument will be widened. Continuum polarization arises in asymmetric scattering, dichroic extinction, and some non-thermal emission processes, for instance. Size distribution and chemical composition of interstellar grains can be estimated by modeling the continuum polarization wavelength dependence  (see, e.g., Rodrigues et al.~\cite{rod97} and Siebenmorgen et al.~\cite{sie14}). Continuum polarimetry can also be used to determine the geometric distribution, sizes, and nature of scatterers in circumstellar environments. Some examples are young stellar objects (Pereyra et al.~\cite{per09}), Be stars (Draper et al.~\cite{dra14}), and binary stars (Rodrigues \& Magalh\~aes~\cite{rod95}). The geometry of supernovae explosions is another example of objects that can be studied using the continuum polarization measurements (Pereyra et al. 2006). The presence of spots in a stellar surface emission can break the symmetry producing a net continuum polarization (Patel et al.~\cite{pat13}). 
A classical example of extragalactic application of continuum spectropolarimetry is the discovery that the polarized flux of Seyfert 2 galaxies can reveal a Seyfert 1 spectrum (Antonucci \& Miller~\cite{ant85}), leading to the unified model of active galactic nuclei. Synchrotron emission from active galactic nuclei can also produce optical continuum polarization (Barres de Almeida et al.~\cite{bar14}). In stellar astrophysics, cyclotron emission in magnetic cataclysmic variables is also a source of continuum polarization (Silva et al.~\cite{sil13}).

\section{Conclusions}
\label{concl}


We demonstrated the ESPaDOnS' ability to measure continuum polarization. A sample composed of HeAeBe stars with known optical polarization and polarimetric standards were used to study the linear polarization. Other objects of different kinds and with expected null circular polarization provide a sample to circular polarization analysis.  We obtained the wavelength dependence of the continuum Stokes parameters for all objects. Data reduction was done using the new OPERA pipeline for ESPaDOnS data. Well sampled continuum linear polarization spectra were measured with polarization levels ranging between 0\,\% and 5\,\% and circular polarization around 0\,\%.

Synthetic broad-band linear polarization computed from the ESPaDOnS spectra was compared with published values to quantify the accuracy of the instrument. The continuum linear polarization measured by ESPaDOnS is fully consistent with the broad-band polarimetry measurements already published. Our tests yielded a polarization degree accuracy around 0.2$-$0.3\,\%. In turn, the accuracy in polarization position angle was better than 5\degr. When continuum circular polarization spectra are analyzed, instrumental polarization of 0.2$-$0.3\,\% is also found for the Stokes \textit{V} parameter. Our results suggest that high quality measurements of the continuum polarization are feasible with ESPaDOnS. 

Many science cases would benefit from continuum polarization measurements. This justifies our study that proves the ability of ESPaDOnS to reliably measure continuum polarization.

Some of the deficiencies presented in Sect.~\ref{instrum} that limit accurate measurements of continuum polarization with ESPaDOnS are minimized in slit-fed low resolution spectropolarimeters (e.g., FORS). For instance, a slit-fed spectrograph does not suffer from random fiber transmission losses due to fiber torsion and misalignment between the star PSF and the fiber entrance. However, as presented in Sect.~\ref{instrum}, we showed that fiber-fed spectrographs can be optimized for continuum polarization measurements by adopting sufficiently long integration times coupled with the frequency of the fiber scrambler, and also by implementing careful guiding to keep the alignment between the target and fiber.


\begin{acknowledgements}
The authors wish to thank to the referee S. Bagnulo for his careful reading. His several comments and suggestions greatly helped to improve the paper. A.~P. thanks FAPESP (grant 13/11473-7) and CNPq (DTI grant 382.585/07-03 associated with the PCI/MCT/ON program). A.~P. is grateful to Douglas Teeple for his support to the OPERA installation. C. V. R. thanks Fapesp (grant 10/01584-8) and CNPq (grant 306103/2012-5). This research used the facilities of the Canadian Astronomy Data Centre operated by the National Research Council of Canada with the support of the Canadian Space Agency. 

\end{acknowledgements}




\end{document}